\documentclass[acmsmall]{acmart}

\AtBeginDocument{%
  \providecommand\BibTeX{{%
    \normalfont B\kern-0.5em{\scshape i\kern-0.25em b}\kern-0.8em\TeX}}}

\setcopyright{acmcopyright}
\copyrightyear{2018}
\acmYear{2018}
\acmDOI{10.1145/1122445.1122456}

\acmJournal{JACM}
\acmVolume{37}
\acmNumber{4}
\acmArticle{111}
\acmMonth{8}

\usepackage{hyperref}
\usepackage{graphicx}
\usepackage{wrapfig}
\usepackage{makecell}
\usepackage{array, booktabs}
\usepackage{caption}
\usepackage{multirow}
\usepackage{multicol}
\usepackage{siunitx}
\usepackage{pbox}

\newcommand{\labeltext}[2]{%
  \@bsphack
  \csname phantomsection\endcsname 
  \def\@currentlabel{#1}{\label{#2}}%
  \@esphack
}
\newcolumntype{C}[1]{>{\centering\arraybackslash}m{#1}}

\begin{document}

\setcopyright{acmcopyright}
\acmJournal{PACMHCI}
\acmYear{2020} \acmVolume{4} \acmNumber{CSCW3} \acmArticle{241} \acmMonth{12} \acmPrice{15.00}\acmDOI{10.1145/3432940}

\title[A System for Interleaving Discussion and Summarization]{A System for Interleaving Discussion and Summarization in Online Collaboration}

\author{Sunny Tian}
\affiliation{\institution{MIT CSAIL}
 \country{USA}}

\author{Amy X. Zhang}
\affiliation{%
  \institution{University of Washington}
  \country{USA}}

\author{David Karger}
\affiliation{\institution{MIT CSAIL}
 \country{USA}}

\renewcommand{\shortauthors}{S. Tian, A. X. Zhang, D. Karger}

\begin{abstract}
  In many instances of online collaboration, ideation and deliberation happen separately from any synthesis of that deliberation into a cohesive consensus document. 
However, this may result in a final document that has little connection to the discussion that came before. 
In this work, we present \textit{interleaved discussion and summarization}, a process where discussion and summarization are woven together in a single space, and collaborators can switch back and forth between discussing ideas and summarizing discussion until it results in a final document that incorporates and references all discussion points.
We implement this process into a tool called  Wikum+ that allows groups working together on a project to create \textit{living summaries}---artifacts that can grow as new collaborators, ideas, and feedback arise, and can shrink as collaborators come to consensus.
We conducted studies where groups of six people collaborated on a proposal using Wikum+ and a proposal using a messaging platform along with Google Docs.
We found that Wikum+'s integration of discussion and summarization helped groups be more organized, allowing for light-weight coordination and iterative improvements throughout the collaboration process.
A second study demonstrated that in larger groups, Wikum+ was more inclusive of all participants and more comprehensive in the final document compared to traditional tools.
\end{abstract}



\maketitle

\section{Introduction}

Cooperation and the sharing of ideas in team production has many benefits, including higher quality output~\cite{beck1993survey}, more innovative ideas~\cite{dyad-prototyping},  increased overall productivity~\cite{dabbish2012social}, and enhanced social relationships~\cite{rice1994describing}. 
Today, online tools provide a liberating medium for people to conduct discussion and collaboration across time and distance.
Even strangers in groups of hundreds or thousands of people can come together to solve open problems and tackle grand challenges~\cite{malone2009harnessing,cranshaw2011polymath,kittur2007he}.

However, collaboration can be difficult, particularly when it is conducted remotely, asynchronously, and at scale.
Online discussion tools such as email, chat applications like Slack, or threaded forums like Reddit, can help groups with conducting remote ideation and deliberation.
However, a major challenge with these tools is refining and combining the mountain of comments into something meaningful~\cite{idea-jam}. 
Comments with important ideas or the outcome of deliberations can be buried deep in or scattered throughout online discussion threads, whereas other comments may be off-topic or redundant~\cite{arkose}. 
Users must sift through repetitive and low quality comments, worsening the creativity of their own contributions and their overall productivity~\cite{siangliulue2015toward}.

As online discussion tools have few synthesis capabilities, sometimes groups use collaborative writing tools such as Google Docs, Dropbox Paper, or MediaWiki to coalesce ideas.
To accommodate discussion on these platforms, some incorporate a separate space for group discussion, such as the Talk Page in MediaWiki or the chat feature in Google Docs. 
However, these spaces are not integrated with the main document, so users cannot reference discussion in the document and vice versa.
Tools like Google Docs also have features for margin commenting, but the comments are primarily used for feedback on already-written content as opposed to initial discussion that builds up to consensus.

\begin{figure}
  \centering
  \includegraphics[width=\columnwidth]{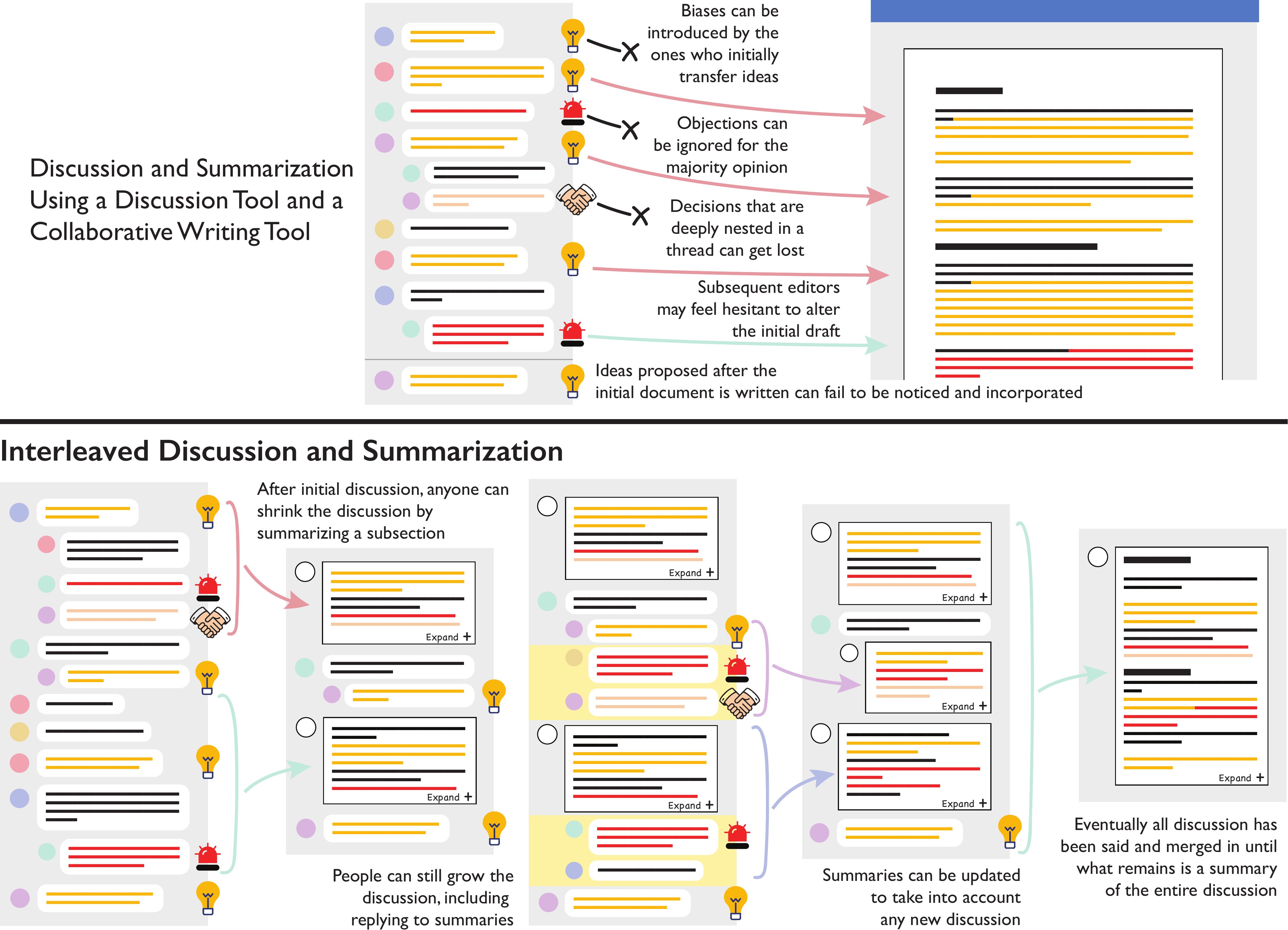}
  \caption[]{Using separate tools for discussion and the synthesis of that discussion into a consensus document can lead to opinions and decisions that are lost in the transfer and an initial draft that becomes difficult to change with new discussion and new contributors. 
  Instead, a system that supports \textit{interleaved} discussion and summarization has summaries that exist in place overlaying the discussion they summarize. This allows discussions to organically continue to grow with new comments, as well as shrink as those comments get merged into summaries, until all perspectives are merged into a single consensus summary that links back to the original comments.}
  \label{fig:interleaved}
\end{figure}

While online discussion tools and collaborative writing tools each have their strengths, there is no single tool that scaffolds all stages of the collaboration process.
Instead, groups oftentimes juggle multiple tools, such as discussing ideas on an online discussion platform before moving to a collaborative writing tool. 
However, this strategy has a number of downsides.
By migrating away from where they had discussion, users lose the ability to directly reference conversation or easily incorporate what people said.
In addition, important ideas and decisions as well as particular perspectives may get lost in the transfer. 
Because the task of reading, transferring, and distilling conversation is hard, it is oftentimes delegated to a particular person, such as a note-taker~\cite{feldman1984development}. This does not scale and also exacerbates issues of bias and lack of comprehensiveness if subsequent writers feel uncomfortable changing the initial draft~\cite{viegas2004studying}.
Finally, new ideas and discussion may come up in the discussion platform while writing has already begun on the writing platform.  As there is no support for synchronizing state between the discussion and document, or tracking progress towards incorporation of the discussion, users have to keep up with both places and continually reconcile their differences.

In this work, we propose an \textit{interleaved discussion and summarization} process to scaffold the entire collaboration life cycle.
Instead of a single discussion stage followed by a single synthesis stage, we represent collaboration as an interleaved process, where ideas can go back and forth between \textit{discussion} periods, when more comments are created, and \textit{summarization} periods, when comments are condensed into summaries (Figure~\ref{fig:interleaved}).
We build this workflow into a tool called  Wikum+\footnote{Wikum+ is an evolution of the existing tool Wikum~\cite{wikum}, a portmanteau of wiki and forum.} that is a hybrid between an online discussion tool and a collaborative writing tool.
Within  Wikum+, users start out having threaded conversations much as in a discussion forum. Users also have tools to curate and condense the discussion, such as by selecting a group of comments and summarizing them. 
Summaries can contain other summaries, allowing ideas and deliberations to eventually be refined into higher-level pieces of writing that encapsulate what was said.
Since new ideas and feedback can come at any point, users also have the ability to reply to summaries or contribute additional comments to discussions that have been already summarized. 
Doing so outdates the original summary; users can then update them to incorporate the new comments, forming \textit{cycles} of summarization and discussion that iteratively refine a summary.

With these features,  Wikum+ allows groups to create a \textit{living summary} artifact that condenses as people come to consensus and refine down ideas, and expands as new users arrive and new ideas get posed.
Different sections can evolve at different paces according to the group's needs---one section may be summarized quickly and stays unchanged, while other sections may need many cycles of discussion and summarization before stabilizing. In addition, users themselves can switch between discussing one facet and summarizing another.

We conducted a lab evaluation to compare how collaboration progresses using Wikum+ versus a control involving a combination of Slack and Google Docs. Thirty-five people in groups of about six were tasked to write a proposal in Wikum+ over the course of three days and then used Slack and Google Docs to write a second proposal, with order counterbalanced.
We found that compared to the control, users in Wikum+ wrote more concisely, had a more organized experience, and could more easily locate both new ideas and topics that still needed work. This allowed users in Wikum+ to make iterative improvements on the summaries and final document. In contrast, groups in the control had trouble transferring ideas from Slack, with one group failing to convert their conversation into a proposal entirely. Interestingly, we noticed some of the groups that had the control condition second replicated strategies they learned from Wikum+ in their Google Doc.

We then performed a second lab evaluation to examine larger groups working together over a longer period of time with staggered entry. Thirty-two people were randomly put into one of two groups, where one group used Wikum+ to write a proposal and the other group used Google Docs to write a proposal.
Unlike the smaller groups with more active collaboration, we found that in the larger, more asynchronous groups, Wikum+ users were significantly more likely than the Google Doc users to find the process to be more inclusive of their ideas, making for a more comprehensive proposal. 
From an independent evaluation of the final proposals, we found that Wikum+ resulted in a more comprehensive, less biased, and more successful final document.

\section{Related Work}

\subsection{Collaboration in Writing}
With the emergence of many collaborative writing technologies \cite{bernstein2010soylent, noel2004empirical}, writing together with a team has grown in popularity for work groups \cite{hackman1987design}, research collaborations \cite{wuchty2007increasing} and student projects in education \cite{kessler2009student}. 
Collaboration in writing has been shown to help authors pool diverse ideas, provide feedback, and write better texts in terms of task fulfilment, grammatical accuracy, and complexity \cite{storch2005collaborative}. Furthermore, authors writing collaboratively generally improve the overall accuracy of their resulting work compared to single-author works \cite{kessler2012collaborative, elola2010collaborative}. 
Convenient feedback and fast response times in these technologies can also increase motivation and creativity of the writers, particularly when generating ideas \cite{lam1995computer}.


An important and well-studied part of this cooperation involves the visibility of edits and contributions in the documents \cite{birnholtz2012tracking, wang2015docuviz}. While increasing visibility of edits can facilitate awareness of changes to a document, studies of collaboration on Wikipedia sites show that making edits completely transparent on documents can result in social friction and inflame conflict, requiring discussion or moderation to draw resolution \cite{kittur2007he,andre2014effects,bryant2005becoming}. But these tools are primarily focused on supporting authors in collectively mutating a document and seeing what changes others have made, rather than discussing whether or not those changes are good ideas. These platforms are not typically used for messaging and communication, making users hesitant to change content written by others in a shared document, not wanting to provoke conflict \cite{birnholtz2012tracking, posner1993people}. This lack of cross-editing change can stifle chances for improvements to the document.

\subsection{Ideation and Creativity Support}
One major benefit of online group work, through collaborative document writing, discussion platforms, or community-based Q\&A websites, is the diversity of ideas a crowd can generate.
By drawing inspiration from peers' ideas \cite{nijstad2002cognitive,siangliulue2015toward,siangliulue2015providing} or external examples \cite{chan2011benefits, lee2010designing, huang2016chordripple}, a community can converge on novel, high quality solutions more quickly than individuals can alone \cite{boudreau2015open}.
However, large groups tend to develop large amounts of shallow and redundant ideas and have trouble deciding on and selecting a good subset~\cite{idea-jam, kittur2009coordination, riedl2010rating}. In particular, with enough people and enough time, groups can achieve abundant levels of divergent thinking (or the generation of ideas in diverse directions)~\cite{guilford1967nature}, but struggle with convergent thinking (or narrowing ideas and identifying the optimal solutions)~\cite{duncker1945problem}---both of which are considered distinct but critical pathways to creativity~\cite{lu2017switching}.

Exposure to repetitive and simple ideas has been shown to hinder creativity because people resort to effort-saving strategies by borrowing from existing ideas~\cite{design-fixation}. On the other hand, seeing high quality, inspirational ideas allows individuals to come up with more diverse and creative ideas, improving overall productivity~\cite{nijstad2002cognitive, siangliulue2015toward}. Examples \cite{sio2015fixation, lee2010designing,chan2015impact}, parallel ideation \cite{dow2010parallel}, and exploration of variations have all been shown to produce higher quality and more novel ideas \cite{nijstad2010dual, rietzschel2007relative}. On the other hand, a poor selection of examples can result in distractions \cite{nijstad2002cognitive} and fixation \cite{design-fixation, kohn2011collaborative}, harming the ideation process. Research suggests that task switching can reduce this fixation and increase creativity \cite{lu2017switching}.
Moreover, highlighting and helping users notice useful, diverse, and creative ideas while hiding or minimizing less useful comments are crucial for both generating and synthesizing the best ideas.


\subsection{Tools for Filtering and Moderating Comments}
Many online collective deliberation systems currently rely on filtering~\cite{lampe2004slash}, voting, a single or few moderators~\cite{discourse}, community flagging \cite{crawford2016flag}, or community rating \cite{lampe2004slash} to pick the best ideas. 
Unfortunately, these methods can introduce biases as well as suppress minority opinions, due to reliance on moderators or under-provision of votes~\cite{gilbert2013widespread}. In particular, having a moderator or a majority vote filter out ideas can oftentimes result in a lack of explanation as to \textit{why} some opinions were not incorporated, besides the fact that it was not favored.
In addition, voting systems have no capability for more sophisticated \textit{refinement} to improve an idea or \textit{synthesis} of multiple ideas. For instance, some competing ideas may be impossible to combine while others become stronger together. Some mechanisms exist for personalized filters, thereby reducing noise, but also increasing the chance of having only one point of view represented \cite{pariser2011filter}. Other tools have been created for automated filtering, detecting spam \cite{mishne2005blocking} and trolls \cite{cheng2015antisocial}.
But these methods all still face problems with inclusion, load-balancing, and effectiveness in sufficiently reducing quantity. 
Furthermore, even after insignificant ideas are filtered out, the challenge of deciding which opinions should prevail in a good but diverse---and potentially controversial---set of ideas still remains.

\subsection{Consensus-Building Discussion Tools}
A number of systems have been developed for online collaborative deliberation that polarize comments by having a voting system or a methodology to determine a ``winning side'' \cite{democracy-os, loomio, kialo, consider-it}. This not only shuts down or hides valid ideas, but also results in a large number of the participants feeling like they've lost something that could have been collaborative.
Furthermore, many consensus-building platforms have little or no affordances for discussion due to the structure and purpose of the system \cite{polis, arguman, quora}. For instance, ConsiderIt places users on a linear scale of agreement with pro-and-con commentary by opinion \cite{consider-it}. Similarly, OpinionSpace allows users to input comments with their ratings of a series of questions, which can then be visualized in a 2-D grid of all inputted opinions \cite{faridani2010opinion}.

Some systems with moderate levels of discourse enforce a great deal of structure when interacting on the interface. Many theories of structured argumentation \cite{maclean1991questions} have manifested into discourse platforms such as Kialo \cite{kialo} and Deliberatorium \cite{klein2011harvest}. On the other hand, Q\&A sites require discussion to be structured in a question-and-answer format \cite{ackerman1990answer}. With limited support for free-form discussion, the types of deliberation that can exist on these interfaces are constricted to ones that result in a ``best answer'' or ``winning side,'' lacking capabilities for many other types of discussions that may result in multiple or a combination of options.

Finally, many platforms allow for more free-form discussion, but either rely on a single summarizer \cite{assembl}, rely on moderators that select representative comments \cite{discourse},  
or rely on polls and surveys based on the discussion to decide on a specific topic \cite{loomio, consider-it}.  
As previously discussed, reliance on these filtering methods can result in biased, non-inclusive, and limited opinions.
By making consensus-building more collaborative, Wikum+ aims to give a greater voice to ideas that may otherwise have been filtered out, voted out, or ignored.

\subsection{Synthesis In Discussion}
One way to handle large discussions is through synthesis. Computational summarization techniques exist for text that are feature-based \cite{gupta2010survey}, cluster-based \cite{jain1999data}, graph-based \cite{erkan2004lexrank}, and knowledge-based \cite{hahn1999knowledge}. But these methods have limitations in dealing with complicated ideas and do not reach the cohesiveness, complexity, or accuracy of human-based synthesis \cite{hahn2016knowledge}. Furthermore, a group may want to choose a subset of the ideas they generate or combine various sets of ideas, requiring more than basic summarization capabilities.

Some efforts, including the original Wikum and Arkose system, have been made to collectively and incrementally distill large quantities of messages or comments into more useful, succinct, and durable summaries \cite{arkose, i-diag, zhang2018making, wikum}. These summaries, when built into the discussion system, allow for a rich synthesis of ideas, encourage reflection \cite{kriplean2012you}, and can help readers explore main topics \cite{wikum} or catch up on missed conversations \cite{zhang2018making}.
But these tools largely focus on completed discussions and do not explore the possibility of continuous and simultaneous conversation in addition to summarization.  Unlike these systems, Wikum+ offers the opportunity to \emph{use summarization to advance the conversation}.  Moreover, the summaries on these platforms are not intended to generate dynamic documents or proposals---where the quality, content, and flow of writing matters more. This leaves the summaries useful only for those interested in the particular page's discussion content, and less so for any greater external purpose, such as sharing the summary document for motivating or influencing change. 

Some systems exist for synthesizing various ideas into more meaningful outcomes. For instance, Climate CoLab allows for the integration of compatible projects into an actionable proposal for fighting climate change \cite{introne2011climate}, while the ``Knowledge Accelerator'' scaffolds small contributions of crowd-sourced information into a single instructive article \cite{hahn2016knowledge}. However, in these systems, once the final proposal or article is created, the discussion and ideas that led up to the resulting document are disregarded and no space exists for continued discussion. New opinions cannot be formed from the previous conversation, and there is no clear way to discuss potential improvements for the document directly on the interface.  
Instead on Wikum+, through interleaved discussion and summarization, users are able to reply within threads linked from summaries or directly to summaries themselves, allowing for continuous rounds of deliberation followed by summary improvements.

\section{Interleaved Discussion and Summarization}

We begin by motivating and describing the components of interleaved discussion and summarization and how it takes and combines elements from both online discussion and collaborative editing tools. In the next section, we describe how it is implemented within the Wikum+ system.

\subsection{Design Goals}

When it comes to design requirements for a collaboration tool, the ability to \textit{synthesize} and \textit{organize} ideas after a period of brainstorming is an essential component.
As mentioned, existing online discussion tools lack sufficient features for making sense of long and unwieldy discussions and redundant comments.
 Echoing prior work on recursive summarization~\cite{wikum}, one way to shrink discussion without simply filtering comments out is to allow users to select subsets of the discussion and summarize them. By doing this in a recursive fashion, users can break up the work of summarizing a large discussion into manageable pieces, while also organizing the discussion into sections. 
 This bottom-up process in addition encourages a \textit{comprehensive} final document, as it informs users what parts of the discussion have been incorporated into the final summary.

However, the recursive summarization approach~\cite{wikum} assumes that a discussion is concluded before summarization begins. In contrast, it is not unusual for new ideas and points of discussion to come up throughout the course of a collaboration, instead of just at the start. This is especially true if a new collaborator joins at a later stage and brings in new insights.
As different aspects of the collaboration task may proceed at different paces, it would be important to support \textit{fluid interleaving} of areas where users are currently discussing versus summarizing within the tool.
In addition, the act of synthesis may itself spur new ideas or objections and refinements to the ideas being summarized. 
Thus it is also important for collaboration tools to support \textit{iterative refinement}, so that discussions that are summarized might be opened for discussion again. In addition, it would be useful for users to easily determine which areas need additional refinement and which have reached consensus.

\begin{figure}
  \captionsetup{singlelinecheck=off}
  \centering
  \includegraphics[width=\columnwidth]{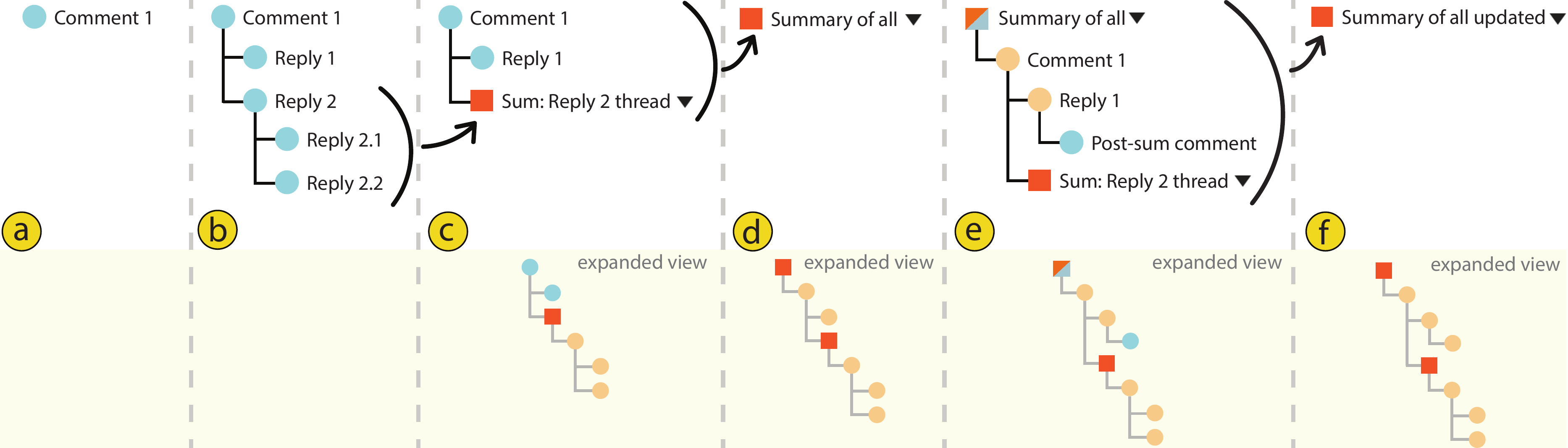}
  \caption[]{ Wikum+ discussions grow with discussion and shrink with summarization. By default (top row), only summary nodes and unsummarized comments are shown, unless summaries are clicked-to-expand (second row). A typical workflow is as follows:
  \begin{itemize}
    	\item[a--b] Users contribute new comments and replies.
        \item[b--c] Users can summarize select comments, adding an orange summary node and changing the blue (unsummarized) comments to yellow (summarized). Collapsed summaries can be expanded to show the content underneath.
        \item[c--d] Summaries can be added at a higher level that include other summaries recursively. The top level summary acts as the final proposal, containing the discussion's contents or outcome.
        \item[d--e] Users can add new comments at any point. Summaries above such comments turn half-blue, half-orange to signify partial summarization of comments below.
        \item[e--f] Users can edit the partial summary to incorporate new comments.
        \item[] Stages d through f loop until the summary reaches a stable state.
    \end{itemize}
    ~\label{fig:progress_viz}}
\end{figure}

A different way to conduct collaboration is wiki-style collaborative editing of a shared document. Tools for collaborative writing provide synthesis and organization capabilities out of the box as users are compelled to make their contributions cohesive to the rest of the document. These tools also allow for iterative refinement of document text through collaborative editing and comments in the margins.
However, there is little scaffolding in collaborative writing tools for the ideation and discussion that typically happen at the outset of a collaboration when plans have not yet been formed.
Initial writers in a document tend to set the tone, giving them a greater say in the conversation and potentially diminishing opposing opinions~\cite{viegas2004studying}.
In addition, while these tools may offer an edit history, it can be difficult to trace where ideas come from and determine whether they are supported by multiple collaborators~\cite{birnholtz2012tracking}.
Thus, a tool that supports \textit{inclusiveness} would give collaborators the opportunity to weigh in before the writing appears to be in a finalized form, as well as scaffold the process so that the work can be spread across more collaborators.


\subsection{Workflow Design}

In Figure~\ref{fig:progress_viz}, we map out the workflow for interleaving discussion and summarization.  
In the beginning, users can leave comments and have back-and-forth discussions, much as they would in a typical forum such as Reddit (Figure~\ref{fig:progress_viz} (a--b)). This format facilitates the free exchange of ideas and opinions without requiring any synthesis.

At any point in the discussion, a user can grab a subset or subthread of the comments and summarize it as a group (Figure~\ref{fig:progress_viz} (b--c)).
One might summarize a thread of discussion because it has converged to a conclusion, or they might summarize a subset of comments that are all variants of the same idea.
Users also have the ability at any point to tag comments, alter their ordering, or move individual comments or entire subthreads to facilitate organization and synthesis.  
As shown in Figure~\ref{fig:progress_viz} (c--d), summaries can also contain other summaries, allowing users to break up the task of summarizing a large discussion into recursive steps. 
Users can access any summarized comments and sub-summaries by expanding their summary.
Steps (b) through (d) replicate the capabilities of recursive summarization studied previously~\cite{wikum}, building up to the top-level summary, which acts as the final document. 
However, unlike previously, users can at any point during these steps leave additional comments to threads or start new threads, allowing for fluid interleaving of commenting and summarizing.

In addition to being able to continue existing comment threads, any user can then add a \textit{post-summary comment} that is in direct reply to a summary or that continues discussing a thread that was previously summarized (Figure~\ref{fig:progress_viz} (d--e)). 
One might choose to leave a comment because they have some feedback about the summary that was written or they feel that the discussion was not actually concluded. They may also be a new collaborator and have new input to provide. 
This ability promotes inclusiveness by allowing users at any point to make contributions to the discussion, even if it has concluded. Leaving a new comment may pose a lower barrier as prior research has shown that users are reluctant to edit or undo writing made by others in wiki-like environments~\cite{wikum,andre2014effects,kittur2007he}.

When comments are added under a summary, the original summary that was written is now no longer comprehensive over the discussion it summarizes and is marked as incomplete.  
At any point, a user can then edit the incomplete summary to incorporate the new comments (Figure~\ref{fig:progress_viz} (e--f)), returning the view to how it was in step (d). Steps (d) through (f) can then loop indefinitely, forming \textit{summarization and discussion cycles} that grow and shrink the discussion, until no one has any new comments to add, all comments are summarized, and users are left with a document that covers the entire discussion. 

With the addition of these later steps, summaries now need not cut off a discussion prematurely but simply act to summarize the current state of the discussion that can then be iteratively refined as new comments come in. 
Because unsummarized discussions and incomplete summaries are marked, users can easily see what areas of the discussion still need to be synthesized.
Given clear markers for what is incomplete, users may also be more comfortable contributing and editing summaries.
In some ways, this portion of the workflow is similar to the process in collaborative writing tools of a user leaving a margin comment on a portion of text in the document and another user coming in to edit the document text and resolve the comment. 
However, this workflow emphasizes that these later comments are not simply minor ``editing suggestions'' but are equal in importance (and deserving of the same rich UI affordances) as the original comments.  It offers comprehensiveness over the discussion by framing the process as incorporating new ideas into a summary as opposed to resolving an issue.

\section{ Wikum+ System}

\begin{figure}
  \centering
  \includegraphics[width=\columnwidth]{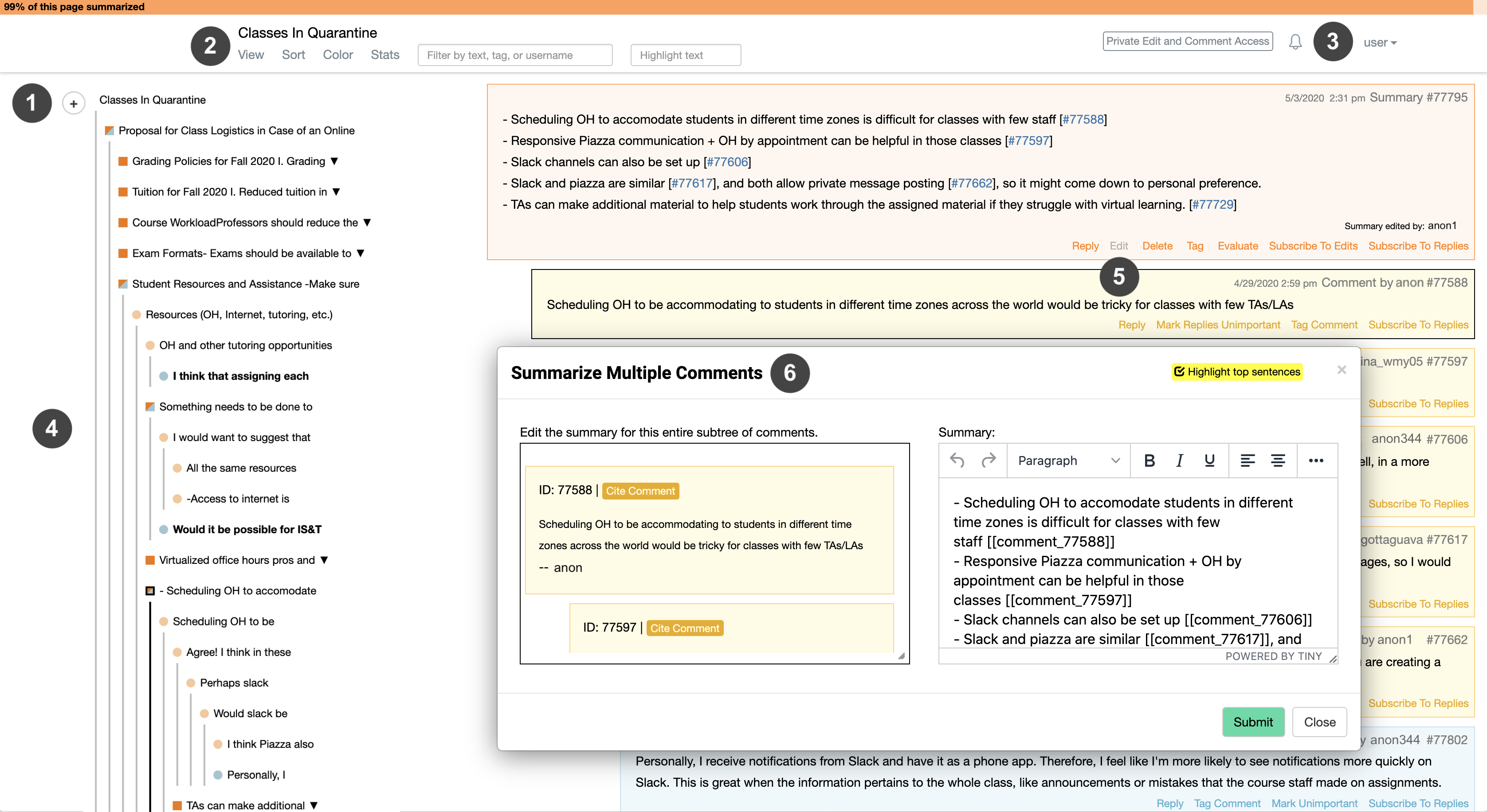}
  \caption{The  Wikum+ UI. (1) New comment button. (2) Drop-down menus. (3) Access Levels and Notifications. (4) Outline View, bold unread indicators; The thread selected in the outline view is displayed on the right.  (5) Locking to prevent race conditions. (6) Summarization modal with content on the left and the summary text in a rich-editing input box on the right.
    ~\label{fig:system}}
\end{figure}

We present the Wikum+ system for supporting collaboration via interleaved discussion and summarization.
 Wikum+ is a website where people can create and share Wikum+ instances\footnote{\textcolor{blue}{\url{ http://wikum.org/}}}.
Figure \ref{fig:system} presents the web interface for an in-progress  Wikum+ instance.
The interface contains an outline view on the left-hand side showing unsummarized comments as blue circles, summarized comments as light orange circles, and summaries as dark orange squares. Indentation and vertical lines are used to show reply relationships as well as summary relationships. 
This view is directly-manipulable, so that users can select, expand and collapse, and move comments and summaries.
On the right-hand side are the full comments and summaries selected by the user from the outline.

\subsection{Collaboration Features}

We describe features of  Wikum+ towards supporting collaboration from initial discussion to the finished final document that encapsulates collaborators' perspectives.

\subsubsection{Features to Support Having and Maintaining a Discussion}

After a user creates a new  Wikum+ instance on the website, they arrive at a blank page where they can begin inviting collaborators and initiating discussion.
The creator of the  Wikum+ instance can set permissions on the instance, such as to give certain individuals commenter-only roles and other individuals editor-only roles (e.g., ability to summarize, move, hide, or tag comments). The creator can also set the entire instance to be publicly commentable, publicly editable, or both (Figure \ref{fig:system} (3)).
Once collaborators arrive, anyone with comment privileges can leave a new comment (Figure \ref{fig:system} (1)) or reply to a comment.

As different groups and stages of collaboration have different cadences, Wikum+ is designed to be usable for a group conducting anything from fully synchronous to fully asynchronous collaboration.
For cases when a group is rapidly discussing or iterating, like they might in a chat environment, Wikum+ supports real-time updates for all actions, including commenting, as well as moving, tagging, flagging, and summarizing comments. 
To prevent potential race conditions, we introduce locks on comments when a user is summarizing them. As Figure \ref{fig:system} (5) shows, editing is disabled for other users while writing a summary. However, other users are still able to simultaneously summarize any other subset of comments or reply to comments that are being summarized. Users are also prevented from moving comments on the outline view when someone else is in the process of moving a comment.

Wikum+ also supports maintaining a discussion when groups are working more asynchronously. 
First, users can subscribe to comments and receive customized email notifications, such as when a user replies to a comment in their comment's reply thread (Figure \ref{fig:system} (3)).
Users can also @-mention other users in comments to notify them.
Finally, we track what comments have been newly added since a user's last visit to the page and bold those comments in the outline view. Clicking on or hovering over the bolded comment removes the ``unread'' marker.

\begin{figure}
    \centering
    \includegraphics[width=\columnwidth]{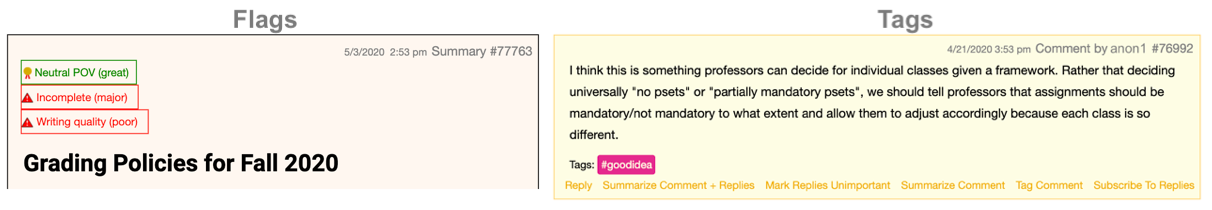}
    \caption{Flags and tags in Wikum+ }~\label{fig:flags_tags}
\end{figure}

\begin{figure}
    \centering
    \includegraphics[width=0.7\columnwidth]{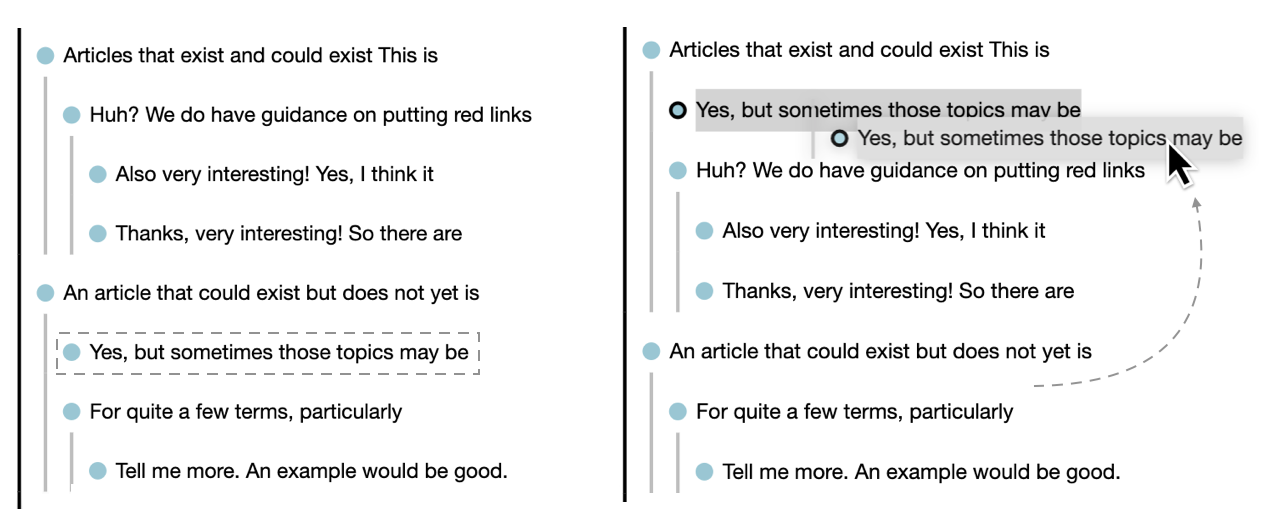}
    \caption{Dragging and dropping a comment from one thread to another.}~\label{fig:drag_drop}
\end{figure}

\subsubsection{Features for Synthesizing and Organizing Discussion}

In order to synthesize discussions, users can summarize a particular subthread of discussion or select group of comments, lower-level summaries, or subthreads, and then summarize them. Users can also choose to summarize a single comment.
Users have access to a rich text editor while editing summaries (Figure \ref{fig:system} (6)). To the left of the editor, they can see the content they are trying to summarize, and can directly \textit{quote} part of the content verbatim or \textit{cite} the text they reference. These references produce a clickable link in the finished summary, allowing readers to jump to the referenced discussion point.
Once a summary is written, it is shown by default in the outline view in lieu of the comments and summaries it summarizes; a user can then click on the summary to expand and show the underlying content. Thus, as more of the content is summarized, the outline view shows fewer items by default.

To aid with collecting similar comments to summarize as a group, users can also add tags to comments (Figure \ref{fig:flags_tags}) and then filter comments by tag. There are additional features for filtering, sorting, and clustering comments and summaries (Figure~\ref{fig:system} (2)). Finally, users can hide comments by marking them as unimportant and delete summaries.  

The stored data model for Wikum+ is the ordered tree presented by default in the left outline view.  Initially, the tree structure is determined by the reply structure in the discussion.   Users modify this tree structure when they create a summary; when they summarize the children of a common parent, the summarized nodes become children of the summary, while the summary becomes a child of that common parent. Users can also drag and drop comments, summaries, and entire threads to different places in the tree.  This permits reording content to improve narrative flow. It also supports the action of gathering related content to be summarized in a single location.  Users can also break the original reply structure when moving comments (Figure \ref{fig:drag_drop}), such as when a comment brings up a tangential point in a discussion thread that is the main topic of a different thread.  For navigation and exploration, the interface also allows users to temporarily sort comments by various attributes (e.g., chronological, by length, etc.) through the drop-down menu (Figure \ref{fig:system} (2)).  However, this sorting is per-user and ephemeral.

\subsubsection{Iterative Refinement of Summaries towards a Comprehensive Document}

Even after a discussion or subthread of discussion has been summarized, the discussion can still continue, and the summary can go through multiple iterations of editing. 
If a user adds a post-summary comment---or a reply to a summary or a subthread of discussion that has already been summarized---the outline view updates the summary icon from a dark orange square to a half-blue, half-orange square, as shown in Figure~\ref{fig:progress_viz} (e). This signals that the summary is now out-of-date due to the newly-added comment. If the newly-added comment is under multiple summaries, due to recursive summarization, all of them become out-of-date.
An out-of-date indicator can also appear on a summary if a user moves a summary, comment, or subthread from elsewhere to under the summary.

Instead of collapsing the content under the summary by default, the outline view expands the content under the out-of-date summary until the unsummarized comment in blue appears. This keeps the view as condensed as possible while not requiring users to dig around to find newly-added comments.
Once a summary is out-of-date, it can be returned back to a dark-orange square if a user edits the summary and checks off that any newly-added comments have now been incorporated. This summary update, incorporating the post-summary discussion, completes a summary-discussion cycle, which can then loop again.

Beyond recursive summarization and the ability for users to edit any summary in a wiki-like way, we also support iterative refinement of summaries by allowing users to evaluate summaries on their neutrality, comprehensiveness, and writing quality. Once a summary has been evaluated, we display flags at the top of the summary (Figure~\ref{fig:flags_tags}). The flags along with the out-of-date summary icons allow users to spot summaries that may need updating and also provide an indication of how to improve the summary. As researchers have found that people are hesitant to edit each other's writing~\cite{wikum}, these signals can also provide encouragement to users to edit summaries.

\subsection{Differences between Wikum+ and Wikum}

 Wikum+ is built on top of the open-source system Wikum~\cite{wikum}.
Within the Wikum system, users could import a concluded conversation that happened elsewhere, such as on Reddit or Disqus, and collaboratively summarize it. As such, the features that previously existed in Wikum include summarization, tagging and filtering, and deletion of comments and summaries. However, Wikum had no support for having a discussion, either at the outset of collaboration, or after summarization had commenced. Thus all the features for having and maintaining discussion are new, as well as all the features for iterative refinement of summaries. Wikum+ also implements real-time updates to all actions, allowing for synchronous, chat-like conversations. Locks help ensure that comments are not lost and summaries are not overwritten, preventing race conditions. Wikum+ also added the affordances for restructuring the discussion by moving subtrees.  We added this feature to assist in grouping content for summarization, as well as in consideration of users who may want summaries to be read in a particular order, much like a collaborative document.  Finally, our outline view is altered from Wikum to allow users to see a snippet of the comment or summary; this presentation more closely mimics outline views in collaborative document software, as well as providing a more informative overview of the discussion.

\subsection{System Implementation}
 Wikum+ is a Django web application and its data is stored in a MySQL database. The front-end is written in JavaScript, HTML, and CSS. We use Django Channels (the Django integration layer), Daphne (the HTTP and Websocket termination server), \texttt{asgiref} (the base ASGI library), and  \texttt{channels\_redis} (the Redis channel layer backend) to handle real-time update functions within a Wikum+ page. The drag-and-drop functionality of Wikum+'s outline view is made possible with the SortableJS library, and the rich text editing feature of summary creation and editing makes use of TinyMCE. The email notifications and notification settings in Wikum+ are accomplished with Pinax Notifications and Pinax Templates. Finally, we use \texttt{jquery-textcomplete} for auto-completion of usernames in the @-user-mentions.

\section{Evaluation}
While Wikum+ permits fluid interleaving of commenting and summarizing, it's unclear what effect this workflow has on the quality of collaboration compared to more traditional online discussion and collaborative writing tools. It is also unclear how groups will use the tool. Will groups resort to discuss-first, summarize-later or make use of the ability to interleave? 

We conducted two lab studies to examine Wikum+ usage compared to a control.
The first study examined smaller groups working more closely on a task, while the second study examined larger groups working more asynchronously on a task. 
Participants for both studies were recruited via university-affiliated mailing lists. We filtered for undergraduate students taking courses remotely as a result of COVID-19 at our university as this experience was relevant to our tasks for groups. 

\subsection{Study 1}

In the first lab study, we focused on the common scenario of a small group of people working together remotely on a task over a period of a few days. We conducted a within-subjects study that compared Wikum+ to a control where groups could use both Slack and Google Docs.
We gathered 6 groups of about 6 people each to collaborate on one proposal in the Wikum+ condition and one proposal in the control condition, with order counterbalanced. We could then compare user experiences with the same group of people in both conditions.

\subsubsection{Participants}
Forty initial participants were split into 6 groups (four groups of 7 people each, two groups of 6 people each). By the end of the study, due to dropouts, 35 participants (Mean age 20.2, 8 Male, 27 Female) had completed the study and the post-study surveys, and the 6 groups ended up consisting of 5 groups of 6 users each and one group of 5 users. Each user that completed the study was compensated \$35.

\subsubsection{Experiment Design}\label{exdes-us1}
This study had two rounds, one for each condition. 
In the first round, each group was tasked with writing a one-page proposal to the university administration on how they would like their university to operate if classes continued to be online in the Fall semester due to COVID-19.
In the second round, each group was asked to write a proposal on how to improve campus dining options.
Details on proposal tasks can be found in Appendix~\ref{appa}.

We randomly assigned participants to one of the 6 groups, with half of the groups (G1--G3) starting with Wikum+ and the other half (G4--G6) starting with the control. Thus at the end of the study, the 6 groups produced 6 Wikum+ proposals and 6 Google Doc proposals. 


\subsubsection{Procedure}
We video-called with each group at the start of both rounds to introduce the round's task. In the first round, group members went through a 5-minute icebreaker. 
During the Wikum+ condition, we added users to a shared private Wikum+ instance, gave a 15-minute tutorial of Wikum+, and told participants that
final proposals should be written in the top level summary in Wikum+.
In the control condition, we added users to a shared private Slack channel and Google Doc.
Groups were able to discuss in their Slack channel or their Google Doc, but their final proposals were to be written in the first page of the Doc. All users were already familiar with Slack and Google Docs.

To spur initial conversation, each group spent 5 minutes at the end of each ``kickoff'' call conversing and ideating synchronously on their given platform(s) about their task. After that, users had 3 days to continue discussion and collectively create a final proposal, thereby completing the task.
After completing each condition, users filled out surveys on their perceived task load \cite{task-load}, their experience, and the document outcome using a Likert scale.

\begin{figure}
    \centering
    \includegraphics[width=1\columnwidth]{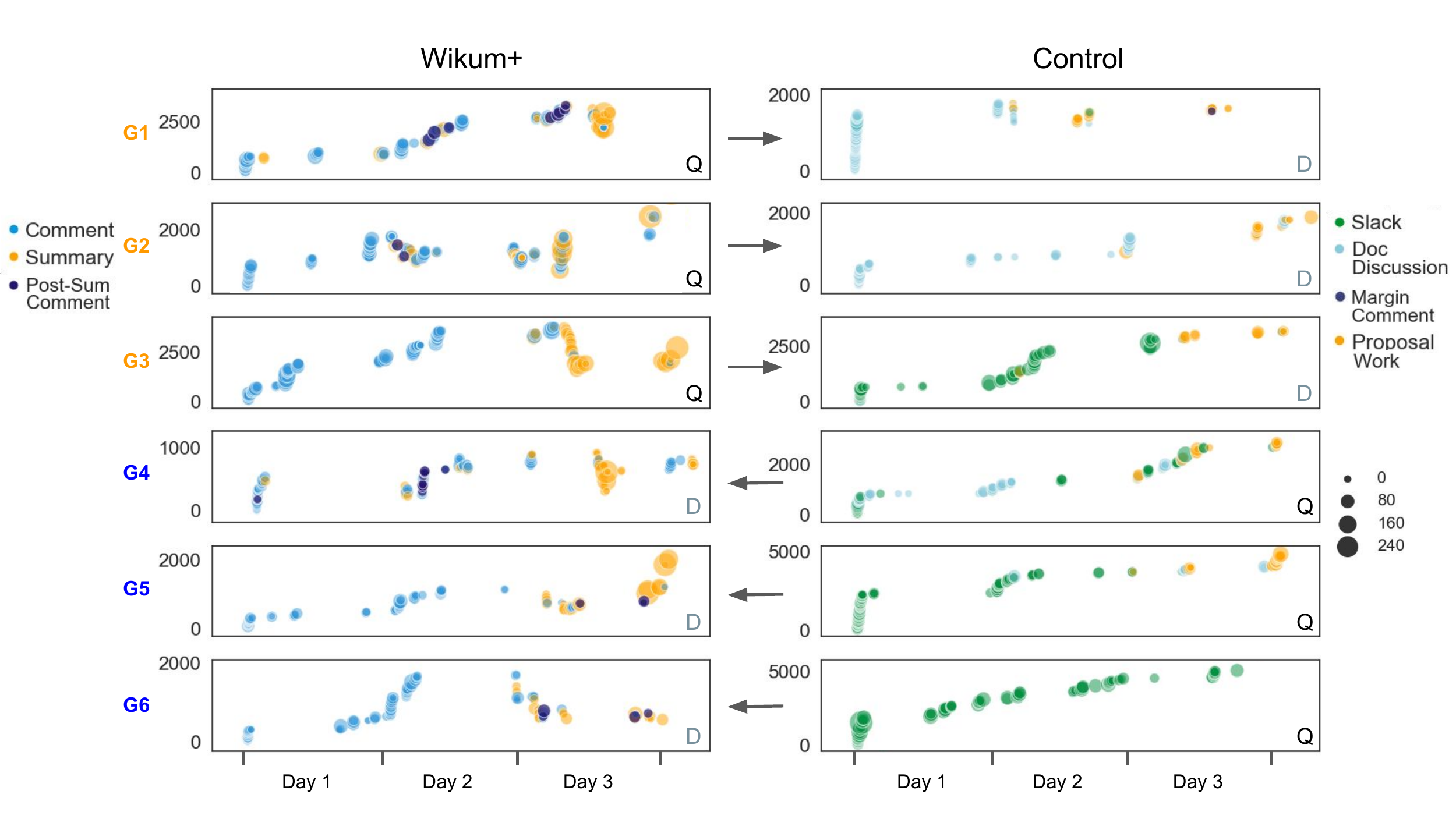}
    \caption{Activity chart over time for each condition (columns) by group (rows). Each node represents a user action.  Each node's size represents the net number of words a user made through an edit. 
    The x-axis is the time of the edit and the y-axis is the number of words shown in the default view at that time.  In Wikum+, this includes all top-level summaries and unsummarized content (because summarized content is collapsed), while in the control, the y-axis counts the total number of words in both Slack and the Google Doc.
    Each graph is labeled with the task topic, either Q (Quarantine Online Classes) or D (Campus Dining Options), and arrows point to which condition happened second. 
}
    \label{fig:us1_all}
\end{figure}

After both rounds had completed, we gave each group one extra day to finish any last-minute edits to either rounds' proposals. This was added after we saw that one of our groups neglected to write up a final proposal in the Google Doc in the control condition, instead only conversing in Slack.
Finally, users answered open-ended questions comparing the two experiences on organization, inclusiveness, and overall success and preference.
The first author then went over the answers to comparative questions to mark whether they stated a preference for one condition versus the other or felt both were equal.

Once all groups completed both tasks, we separately recruited 18 new people by messaging various undergraduates at the same university as the users to rate the quality of proposals written by each group in each condition. Each rater was given one proposal from a Wikum+ condition and one proposal from a control condition, with both proposals on the same topic (and thus written by different groups). We asked raters to rate the argument quality, comprehensiveness, level of bias, and writing quality of each proposal, and also provide their assessment of which proposal was better overall. In this way, we evaluated both the process of collaboration as well as the product.

\subsubsection{Results}
\textbf{ Wikum+ users made use of interleaved discussion and summarization while control users split discussion and summarization.}
Figure~\ref{fig:us1_all} shows the activity in each group over time in both the Wikum+ condition (left) and the control condition (right). Note that G1, G2, and G3 had the Wikum+ condition first, while the other groups had it second.
We note that engagement is higher on the first task topic (online education due to COVID-19), irrespective of condition; this is likely due to high interest among the student population regarding the topic, as well as dropout by 3 users in total from Round 1 to Round 2 (one user each in G1, G4, and G6). 
Users stated that the Round 1 topic was ``\textit{more pressing}'', resulting in ``\textit{more ideas, more thoughts, and more solutions}.'' 

\begin{table}[]
    \centering
    \footnotesize
    \setlength{\tabcolsep}{0.45em}
    \begin{tabular}{|C{3.2cm}|c|c|c|c|c|c|c|c|c|c|c|c|}
        \hline
         &  \multicolumn{4}{c|}{\textbf{G1}} & \multicolumn{4}{c|}{\textbf{G2}} &  \multicolumn{4}{c|}{\textbf{G3}} \\
         & \multicolumn{2}{c|}{\textbf{W+}} & \multicolumn{2}{c|}{\textbf{C}}   & \multicolumn{2}{c|}{\textbf{W+}} & \multicolumn{2}{c|}{\textbf{C}} & \multicolumn{2}{c}{\textbf{W+}} & \multicolumn{2}{c|}{\textbf{C}} \\
         & Com & Sum & Slack & Doc  & Com & Sum & Slack & Doc  & Com & Sum & Slack & Doc \\
        \hline
        Avg. Words Added Per User & 694.8 & 421.9 & 0 & 360.3 & 638 & 342.5 & 0 & 316.3 & 690 & 455.5 & 447.3 & 112.5 \\
        \hline
         Avg. Words Added Per User (Total) & \multicolumn{2}{c|}{1116.7} & \multicolumn{2}{c|}{360.3} & \multicolumn{2}{c|}{980.5} & \multicolumn{2}{c|}{316.3} & \multicolumn{2}{c|}{1145.5} & \multicolumn{2}{c|}{559.8}\\
        \hline
        Total Words In Proposal & \multicolumn{2}{c|}{717} & \multicolumn{2}{c|}{360} & \multicolumn{2}{c|}{363} & \multicolumn{2}{c|}{469} & \multicolumn{2}{c|}{657} & \multicolumn{2}{c|}{607}\\
        \hline
        \hline
        &  \multicolumn{4}{c|}{\textbf{G4}} &  \multicolumn{4}{c|}{\textbf{G5}} &  \multicolumn{4}{c|}{\textbf{G6}} \\
         & \multicolumn{2}{c|}{\textbf{W+}} & \multicolumn{2}{c|}{\textbf{C}}   & \multicolumn{2}{c|}{\textbf{W+}} & \multicolumn{2}{c|}{\textbf{C}} & \multicolumn{2}{c}{\textbf{W+}} & \multicolumn{2}{c|}{\textbf{C}} \\
         & Com & Sum & Slack & Doc  & Com & Sum & Slack & Doc  & Com & Sum & Slack & Doc \\
        \hline
        Avg. Words Added Per User & 317.3 & 148.7 & 153.7 & 260.3 & 230.3 & 338 & 609.2 & 194.8 & 359.9 & 92.8 & 725 & 25.4 \\
        \hline
        Avg. Words Added Per User (Total) & \multicolumn{2}{c|}{466} & \multicolumn{2}{c|}{414} & \multicolumn{2}{c|}{568.3} & \multicolumn{2}{c|}{804} & \multicolumn{2}{c|}{452.7} & \multicolumn{2}{c|}{750.4}\\
        \hline
        Total Words In Proposal & \multicolumn{2}{c|}{229} & \multicolumn{2}{c|}{656} & \multicolumn{2}{c|}{758} & \multicolumn{2}{c|}{891} & \multicolumn{2}{c|}{205} & \multicolumn{2}{c|}{175}\\
        \hline
    \end{tabular}
    {\renewcommand{\arraystretch}{1.3}
    \begin{tabular}{|c | C{1cm} | C{1cm} |}
    \multicolumn{3}{c}{\vspace{-1mm}}\\
        \hline
         & \textbf{W+} & \textbf{C} \\
        \hline
         Avg. Words Added Per User (Total) & 788.3 & 534.1\\
        \hline
        Avg. Words In Proposal  & 488.2 & 526.3\\
        \hline
    \end{tabular}}
    \caption{Average user contribution levels by group in Wikum+ and the control. The top row (G1, G2, G3) had the Wikum+ condition followed by the control. The bottom row (G4, G5, G6) had the control condition followed by Wikum+. Note that 3 users dropped out before or during Round 2 and that Round 1 received more engagement.}
    \label{tab:contributions}
\end{table}

As can be seen in Figure~\ref{fig:us1_all}, in the Wikum+ condition, groups began by primarily contributing comments, which increased the amount of displayed text. This was followed by a period of interleaved summarization and commenting, where overall discussion size shrank and grew. Finally, groups ended with primarily editing summaries towards writing a final proposal in a top-level summary node; in some cases like G6, this meant shrinking large summaries and discussions down, while in other cases like G5, it meant adding new text to the final summary to make it a cohesive document. 
We specifically note the instances where users ``re-opened'' an existing summary for discussion via the \textsf{Post-Sum Comment} markers in dark blue.

In the control condition, there was greater variety in behavior. Users made use of Slack, the Google Doc document itself, and marginal comments in the Google Doc to have discussions.
Because we asked Google Doc users to leave their proposal in the first few pages of the document,
we can separate out edits in the Google Doc document that are proposal edits (\textsf{Proposal Work} in Figure~\ref{fig:us1_all}) from those that are contributing discussion (\textsf{Doc Discussion} in Figure~\ref{fig:us1_all}).
The first author went through each of the Google Docs and their edit history manually to verify that this occurred.

In this study, only G1 made use of marginal comments at all, and only one of those comments was directly on the proposal content.
Groups 1 and 2, both of whom had the control condition second, chose not to use Slack at all, working entirely in their Google Docs (Table~\ref{tab:contributions}).
Group 6, on the other hand, who started out in the control condition, worked solely in Slack for the entire time period and did not add anything to their Doc. After the 3 days were already passed, one member went back to write a summary of the Slack conversation.  
Overall, no groups had significant interleaving of discussion versus proposal writing in the control; instead, groups such as G3 worked in two stages, one involving discussion on Slack or Google Docs, followed by editing the Google Doc with the final proposal.  




\textbf{Round 1 Wikum+ groups reused strategies from Wikum+ when working in the Google Doc in Round 2.} We noticed that users in Google Docs drew inspiration from the practices that were beneficial in Wikum+ when working in the control condition during Round 2. In particular, both Groups 1 and 2 used bullet points to create threaded discussions in the Google Doc, simulating the nested discussion tree in Wikum+. A Group 1 user explained that they were ``\textit{immediately overwhelmed by the amount of text}'' in the Doc, and some users in the group started removing discussion text and replacing it with summarizing text, much as they did in Wikum+, though Wikum+ preserves the original discussion.
As one user started summarizing and working on the Doc proposal, they also moved entire portions of the discussion that they had summarized into the proposal (Figure~\ref{fig:us1_all}, G1). Another user avoided deleting the discussion text but struck out text that they had incorporated. 
While deleting or striking out text reduces clutter and may help users see the remaining discussion that still needs to be included in the proposal, without a platform that explicitly supports bottom-up summarization, users lose the direct links pointing from the proposal text to the discussion that generated and inspired it.

While Group 3 also had the control condition for the second round, unlike Groups 1 and 2, they still discussed in Slack. Right at the beginning of Day 3 in Figure~\ref{fig:us1_all}, there is a large green dot representing one user who took the time to go through all of their group's Slack conversation, summarize each Slack reply thread into a bullet-point list, and send it to the group. Afterwards, other users built off of this Slack message to write up the proposal in the Doc. Each bullet point in the list summarized one discussion thread in the Slack, much like a summary would in Wikum+, but just like in the Doc from Groups 1 and 2, this lacks a direct reference to the conversation thread that sourced the summary. Furthermore, this single-person summarization process requires substantial effort and is not scalable to larger conversations. 

In comparison, on Wikum+, users' summaries were situated above the comments and summaries they summarized, thereby keeping the original thought process and deliberation intact. Users also used citation features, creating a reference that allows others to see \textit{why} some portion of the summary was incorporated.  

\begin{figure}
    \centering
    \includegraphics[width=1\columnwidth]{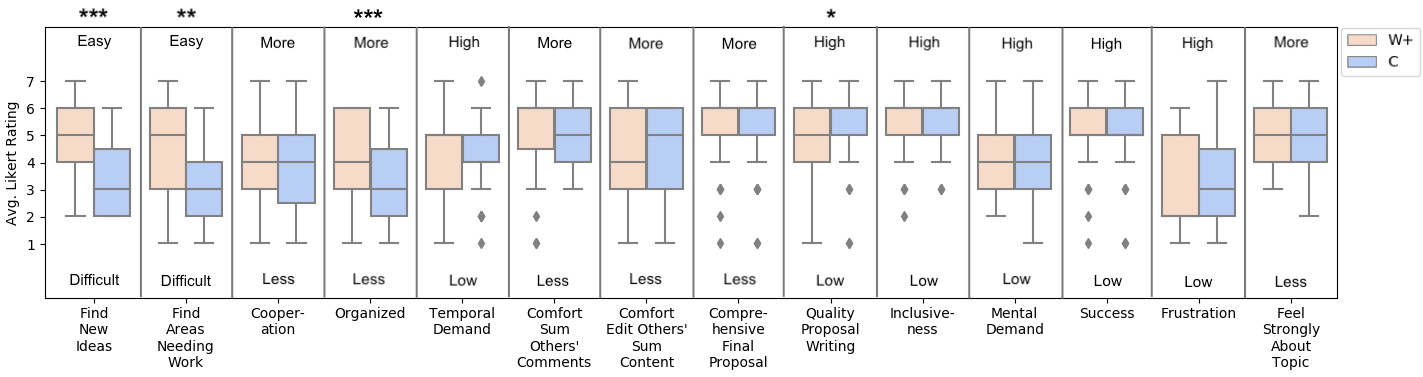}
    \caption{Box plots showing metric values (out of 7), by condition, from the conditional post-study surveys. Metrics that resulted in significantly different means from paired sample $t$-tests are starred ($p<0.05$*;  $p<0.01$**; $p<0.005$***).}
    \label{fig:all_metrics}
\end{figure}

\textbf{Users in Wikum+ found it easier to locate both new ideas and topics that still needed work, allowing for iterative improvements on summaries.} Many users in the post-study survey pointed out that a number of Wikum+ features helped them find areas in the conversation to jump into. When users came back to the Wikum+ page after a period of time, bold unread markers and blue comments under summaries helped them find new ideas introduced while they were gone. Users also reported that summary flags, the summarization progress bar, blue unsummarized comments, and partial summary nodes helped them easily pinpoint areas that still needed work.
From the survey metric results, users felt that Wikum+ helped
significantly in discovering newly introduced ideas ($t=4.30$, $p=0.00013$, Figure~\ref{fig:all_metrics}) and in locating topics that had not been included or still required further deliberation ($t=2.92$, $p=0.0063$, Figure~\ref{fig:all_metrics}).

\begin{table}
\centering
\begin{minipage}{\columnwidth}
    \centering
    \begin{tabular}{c|C{2.1cm}|C{2.3cm}|C{2.8cm}|C{2.8cm}} 
      \Xhline{2\arrayrulewidth}
      & \textbf{\small{Preferred System}} & \textbf{\small{More Organized}} & \textbf{\small{More Inclusive, Cooperative}} & \textbf{\small{More Successful Proposal}}\\
      \Xhline{2\arrayrulewidth}
      Wikum+ &  74.3\% &  85.7\% &  65.7\% &  54.3\% \\
      \hline
      Slack and Doc &  25.7\% &  14.3\% &  17.1\% &  34.3\% \\
      \hline
      Equal &  0 &  0 &  17.1\% &  11.4\%
    \end{tabular}
\caption{Qualitative Comparison Survey Results}~\label{tab:qual-compare}
\end{minipage}
\end{table}

A crucial difference between the two conditions was the iterative improvements made to Wikum+ summaries that were later reflected in the final proposal.
In Wikum+, 5 out of the 6 groups added new replies within existing summaries---or \textit{post-summary comments}---that were all addressed or incorporated either through edits or evaluations of the summary.
Three of the groups had post-summary comments that contended with or made suggestions to the summaries they were created under, resulting in improvements and edits to all of those summaries. Other post-summary comments continued part of a discussion already incorporated in the summary, voiced agreement to the summary they replied to, provided additional details and personal anecdotes, or proposed solutions to problems listed in previous comments. These post-summary discussions resulted in more details, solutions, and improvements in the existing summaries.


In the control condition, only one user, who worked in Google Docs for Round 2 after working in Wikum+ for Round 1, made a single marginal comment directly on the summary content, reminding the others on the task objectives. This led to some restructuring of the proposal, and the comment was marked as resolved. Apart from this one user, no one else made use of marginal comments in the Doc, and many expressed that they had trouble finding newly added content or areas that still needed work.
One user explained ``\textit{re-reading over the entire discussion was the only way to determine new content... [and] what had or hadn't already been added,}'' as there was almost ``\textit{nothing to help track unincorporated points of discussion.}'' A handful of users resorted to using the edit history to find recent changes. One user said: ``\textit{I really had to think about how each of the new updates would contribute to the overall task---it was a little overwhelming.}'' This lack of situational awareness and decreased use of post-summary comments resulted in far fewer opportunities for iterative refinement in the final document.

\textbf{Although users wrote more text in Wikum+, when it came to the proposal, Wikum+ users wrote concisely and directly, while Google Docs users introduced greater narration.} On average, users wrote around 254 more words in Wikum+, including both discussion and proposal text, than in the control (Table \ref{tab:contributions}).
Paired sample $t$-test analyses comparing the number of words added by user in the Wikum+ condition against the number of words added by user in the control yield significant differences, providing evidence that a greater quantity of contributions occurred in Wikum+ ($t=2.31$, $p=0.027$).

However, on average, Doc proposals had about 38 more words than Wikum+ proposals (Table \ref{tab:contributions}), and were written with what users deemed a higher quality of writing ($t=-2.26$, $p=0.03$, Figure~\ref{fig:all_metrics}). This suggests that a significant portion of the contributions in Wikum+ were made prior to creating the proposal and that the users were spending more time on gradually synthesizing the discussion along the way. This is supported by the fact that over 20\% of the text added in each group's Wikum+ page consisted of summary text (Table \ref{tab:contributions}).

We found that the proposals in Wikum+, written in top level summary nodes, were organized by the summaries created for each topic thread. Many of the Wikum+ proposals also maintained the bullet-point structures found from the groups' prior summaries, making them more concise. In contrast, users had only the conversation to go off of in most groups during the control condition, the exception being Group 3, where one user synthesized the entire discussion into a summary before the group wrote the proposal. It is possible that this lack of pre-existing summaries in the control resulted in greater narration in the Doc proposals, and that the succinct style of intermediate summaries in Wikum+ influenced the writing style of the following proposals.

\textbf{Users in Wikum+ found that the writing process was more organized than in the control.} 
From the post-study surveys, we found that to many users, Wikum+'s process provided their group structure, allowing for gradual, but visible, progress. A paired sample $t$-test yielded significant evidence that users felt Wikum+ provides greater organization ($t=3.23$, $p=0.003$, Figure~\ref{fig:all_metrics}) and in the comparison survey, 85.7\% of users felt that their experience on Wikum+ was more organized (Table~\ref{tab:qual-compare}).

One user explained that the interface of Wikum+ ``\textit{helped coagulate the comments so that none are lost in the cracks... As you summarize, the comments are displayed beside the summary, so you can ensure you are touching on all topics discussed.}'' 
Another user said they ``\textit{liked building up the summaries in small contributions}'', making it ''\textit{easier when bringing [the discussion]... into the final document.}'' Wikum+ users also utilized the drag-and-drop feature to ensure that comments fell under the right threads, aiding their synthesis process.
In contrast, digging through comments in Slack and then collecting the ideas in Google Docs was harder to organize. For instance, the users in the Group 6 control neglected to write a proposal and instead chatted in Slack during the given three days. Members explained that their ideas were scattered and disorganized in the Slack channel. As one user put it, ``\textit{there was an initial activation barrier to overcome}'' in collecting the thoughts and moving them over to the Doc, for which no one took initiative. 

\begin{table}
\centering
\begin{minipage}{\columnwidth}
    \centering
    \begin{tabular}{C{0.7cm}|C{2.3cm}|C{2.4cm}|C{2.4cm}|C{2.2cm}|C{1.4cm}}
      \Xhline{2\arrayrulewidth}
      & \textbf{\footnotesize{Avg. Argument Quality (1: low, 7: high)}} & \textbf{\footnotesize{Avg. Comprehensiveness (1: low, 7: high)}} & \textbf{\footnotesize{Avg. Level of Bias (1: neutral, 7: biased)}} & \textbf{\footnotesize{Avg. Writing Quality (1: low, 7: high)}} & \textbf{\footnotesize{More successful}}\\
      \Xhline{2\arrayrulewidth}
      \footnotesize{W+} &  3.28 &  3.72 & 4.56 & 3.17 & 9 \\
      \hline
      \footnotesize{C} & 3.33 & 3.67 & 4.39 & 3.39 & 8 \\
    \end{tabular}
\caption{Evaluation of Proposals. 18 non-participants scored a pair of proposals under the same task, one from each condition so that each proposal was scored by 3 raters. One rater found the two proposals equally successful at accomplishing the task so they are omitted from the final column.}~\label{tab:prop-eval}
\end{minipage}
\end{table}

\textbf{Users found their process on Wikum+ more cooperative and inclusive than in the control.} 65.7\% of users felt that Wikum+ was more inclusive of ideas and allowed for greater levels of cooperation, with the remaining proportion split between the control experience and finding the two equally inclusive (Table~\ref{tab:qual-compare}). Although not statistically significant, users also indicated that they experienced higher levels of cooperation (Average Cooperation -- W+ : 4.4, C: 4; $t=1.14$, $p=0.26$) and inclusion (Average Inclusion -- W+ : 5.6, C: 5.3; $t=0.98$, $p=0.33$) in Wikum+ compared to the control.
A user explained that in Wikum+, ``\textit{being able to summarize helped people interact with others' ideas and not just focus on their own,}'' broadening perspectives, while still ``\textit{[being] able to add your own ideas into the summaries.}'' This made ``\textit{the summary mechanic feel like a very collaborative activity}.'' 

\textbf{Quality of the overall experience and final proposals.}
From the comparison survey, users expressed that more so than the final product, their experience of collaborating in the Wikum+ condition exceeded their experience in the control in terms of organization and inclusion (Table~\ref{tab:qual-compare}). In particular, 74.3\% of users preferred Wikum+ over Google Doc and Slack as a system. 

In terms of the resulting document, 54.3\% of the participants found the Wikum+ proposal more successful compared to the 34.3\% of users who found the Doc proposal more successful (Table~\ref{tab:qual-compare}). The remaining users found the two proposals about equally successful.

However, from the evaluation of proposals conducted by independent raters, the raters scored the proposals similarly across the board for average levels of argument quality, comprehensiveness, bias, writing quality, and success (Table \ref{tab:prop-eval}). Raters were also split when asked which proposal was more successful overall.


\subsection{Study 2}

In the second lab study, we turn towards the common scenario where a larger group of people are asynchronously working on a shared task remotely over a longer period of time.
In many cases today, a user might start a Google Doc and share it with a wide audience, such as by posting it on social media, with a prompt asking for open contributions. Participants may not be in a pre-existing workteam or close collaborators with each other.
For this reason, we compare the use of Wikum+ against a control involving just Google Docs,  removing the Slack component.


\subsubsection{Experiment Design}
We recruited 40 participants using the same methods as Study 1. Participants were randomly assigned into two groups: one for Wikum+ and one for a Google Doc. 
Due to dropouts, in the end, 15 users contributed to Wikum+ and 17 users contributed to the Doc at least once during the course of the study. After the 5 days, all Wikum+ participants and 14 Doc participants completed the post-study survey. The results below are based on what took place during the study with all 32 participants (Mean age 20.0, 8 Male, 24 Female), as well as the experiences reported by the 29 participants who completed the survey (Mean age 20.0, 6 Male, 23 Female). Each user that completed the study and survey was compensated \$15.

\subsubsection{Procedure}
In this study, the control group was only given a Google Doc to keep each group on a single platform. 
The Quarantine Online Classes task (Appendix \ref{appa}) was given to both the Wikum+ group and the Google Doc group due to higher levels of engagement in Study 1. We created an 8-minute video tutorial and an online written tutorial for the Wikum+ users to review before starting on their task.

We staggered sending out the initial instruction emails to users over the course of 6 hours with the intention to also stagger the starting times of the users. This way, some users would arrive after some discussion had already occurred, allowing for varying degrees of catching up and the introduction of new ideas along the way. Both groups were given the same deadline, giving each user between 5 to 5.5 days to work asynchronously on their assigned interface.

The Wikum+ group was instructed to write their proposal in a top-level summary node and the control group was instructed to write their  proposal in the first few pages of the Doc. After the deadline, each user filled out a post-study survey based on their condition, similar to the surveys from Study 1. 

After both groups completed the task, we separately recruited 6 new undergraduate students to evaluate and compare both proposals on the same evaluation metrics from Study 1.

\subsubsection{Results}
\begin{figure*}
  \centering
  \includegraphics[width=1\columnwidth]{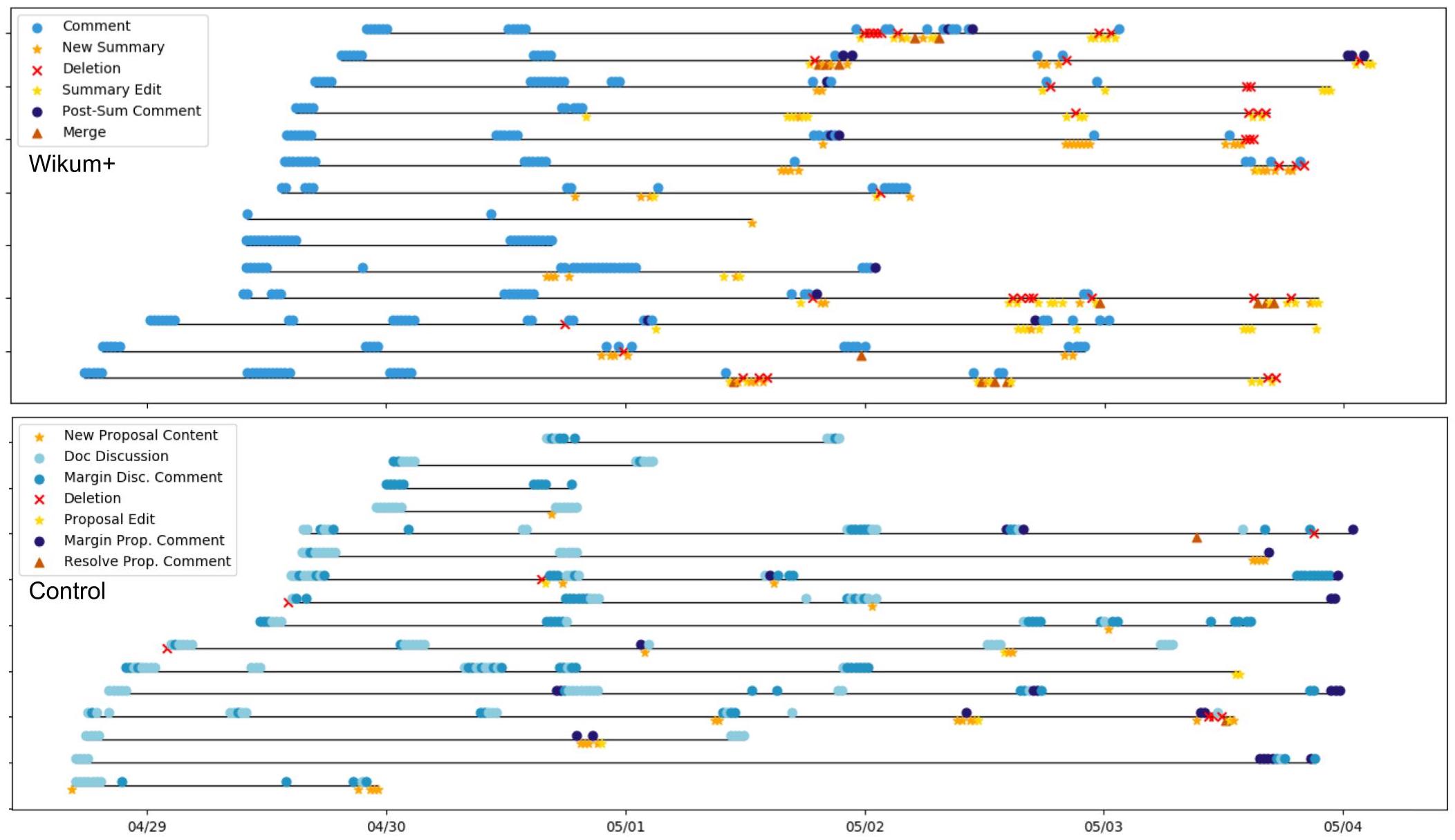}
  \caption{
  Timelines of user contributions in the two conditions. 
  Each line represents one user and markers symbolize the type of contribution made and the x-axis shows absolute time. Two outlier users are omitted, one from each group, who both started after Day 3 of the study and made $\le 5$ total contributions. Note that no margin discussion comments were resolved in the control condition. While some users started working on the proposal earlier in the control, a greater proportion of users contributed substantially to summaries and proposal content in Wikum+. A lack of organization and inclusion resulted in fewer proposal contributions by many members in the control.}~\label{fig:user_timelines}
\end{figure*}

\textbf{Compared to participants in Study 1, users in the control group increased marginal comment usage while users in Wikum+ continued to interleave discussion and summarization, allowing for summary refinements.}
In this study, the control group made heavy use of the marginal commenting functionality in Google Docs, allowing for the group to remark directly on the proposal content and find areas that still needed work (Table~\ref{tab:large_group_contributions}). We saw an increased use of this commenting feature as time progressed and changes in the Doc itself became harder to detect (Figure~\ref{fig:user_timelines}). Google Doc comments also allow for email notifications, so users could know when and where new comments were added. Likely due to the asynchronous nature of the study, participants did not use the Doc chat feature.

\begin{figure*}
  \centering
  \includegraphics[width=1\columnwidth]{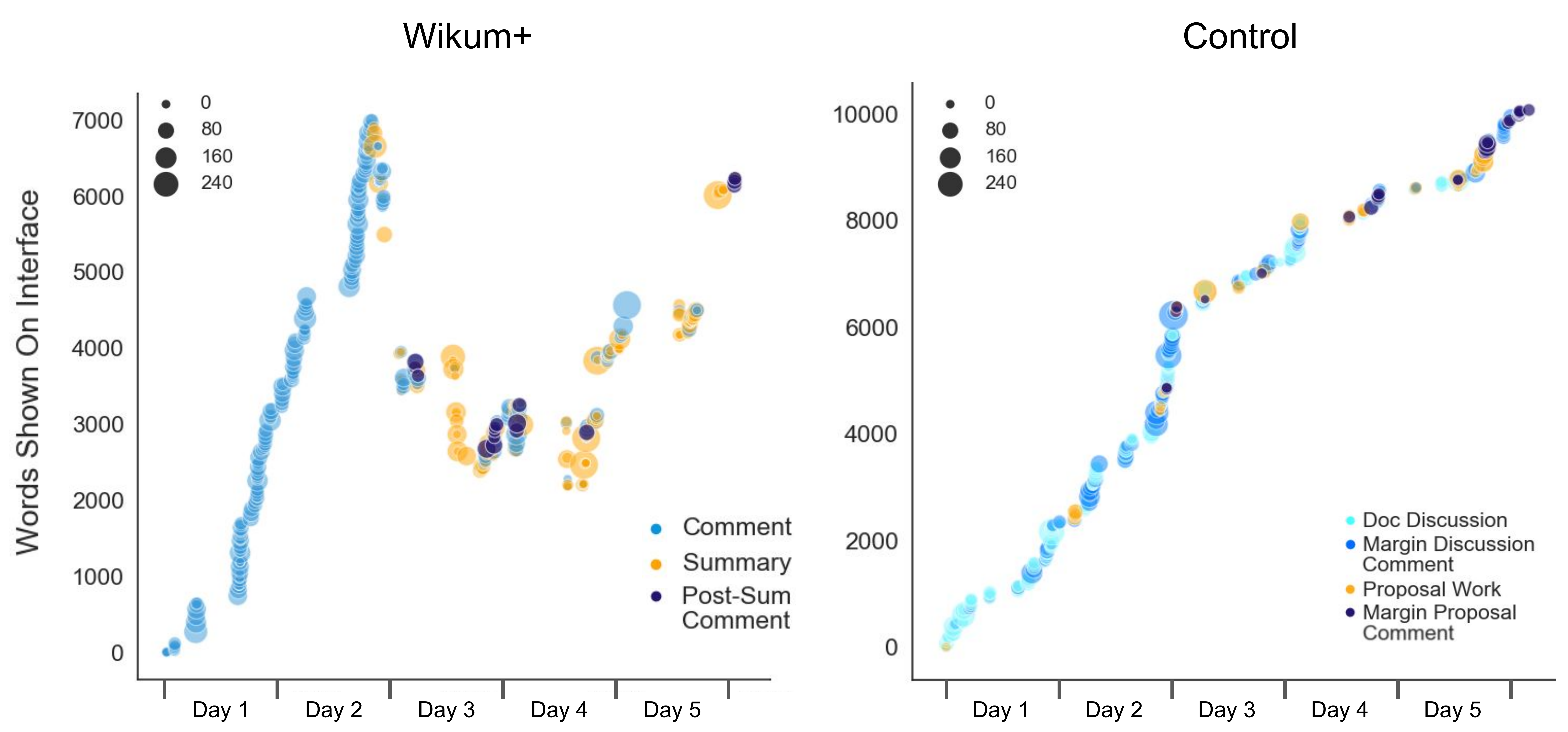}
  \caption{
  Plot of user activity in the two conditions over time. 
  Each node is a user's comment or edit to a summary/proposal, with the size equal to the number of words. The y-axis represents the number of words shown by default on each interface. In Wikum+, this value includes all top-level summaries and unsummarized comments (because summarized comments are collapsed), while in the control, the y-axis counts the number of words in the Google Doc.}~\label{fig:large_group_overall}
\end{figure*}

In Figure~\ref{fig:large_group_overall}, we can see a contrast between the events of each condition. After around two days of discussion, Wikum+ users began summarizing rapidly, synthesizing and condensing many of the page's threads and subthreads. However, discussion continued and some users voiced disagreement with some of the points made in the summaries, resulting in both edits to the existing summaries and new summaries created, incorporating the divergent ideas. In the final day, Wikum+ users primarily edited and combined existing summaries into a final top-level proposal. In comparison, Doc users spent a large proportion of their time in disagreement, trying to determine which opinions to include in the final proposal. In the end, the group did not finish writing their proposal, as some of the requirements from the task were incomplete, and many comments left by users went unaddressed.

\begin{table} \small
\begin{tabular}{|c C{2.5cm}|C{2.7cm}|C{2.8cm}|C{2.7cm}|}
\hline
Condition & Type & Avg. Words Added & Avg. Words Added (Total) & Words In Proposal \\
\hline
\multirow{2}{*}{W+} & \footnotesize{Comment} & 714.8 & \multirow{2}{*}{1151.7} & \multirow{2}{*}{1403} \\
\cline{2-3}
   & \footnotesize{Summary} & 436.9 & & \\
\Xhline{2\arrayrulewidth}
\multirow{3}{*}{C} & \footnotesize{Doc Discussion} & 238.6 & \multirow{3}{*}{608.5} & \multirow{3}{*}{960} \\
\cline{2-3}
   & \footnotesize{Margin Comment} & 305.8 & & \\
\cline{2-3}
   & \footnotesize{Proposal Work} & 64.1 & & \\
\hline
\end{tabular}
\caption{Average user contribution levels in both the Wikum+ and the control conditions.}~\label{tab:large_group_contributions}
\end{table}

\textbf{Although both conditions had comments disagreeing with summary content, Doc users decided to vote for a majority voice, while Wikum+ users iterated on the summaries to include multiple opinions.} 
The first subtopic on mandatory versus optional Pass or No Record grades in the Task caused divisions in both groups, but each group handled this disagreement differently. In the Doc, a couple of users took charge in setting up a system where each user could summarize their viewpoints in the proposal. Then, users could sign off on which idea they were in favor of, only keeping the majority-vote summary in the final proposal.

As discussed previously, the under-provision of votes and a reliance on moderators can lead to biases and suppress minority ideas~\cite{gilbert2013widespread}. Thus, one of our goals with Wikum+ was to promote minority voices and encourage conversation, instead of only voting, to reach consensus. Users in the Wikum+ condition decided that instead of choosing to only express the majority opinion, they would update summaries to include the other opinions, allowing for all voices to be heard.


In Wikum+,  we saw 11 areas in the discussion where comments were created that contended with or made suggestions to summaries (Figure~\ref{fig:user_timelines}). Of these post-summary discussions, 8 led to improvements in the final proposal. Similar to in Study 1, these Wikum+ post-summary comments fell into a few major categories: adding details or personal examples to existing content, ideating solutions to previously listed problems, asking a summarizer to add more details, questioning or discussing the effectiveness of a proposed solution, and answering questions and issues posed in previous comments.
In contrast, we saw 8 marginal comment threads in the Doc disagreeing with or suggesting changes to the proposal content. As in Wikum+, some of these post-summary comments asked for more details in the proposal and questioned the effectiveness of a suggested solution. However, other post-summary comments voted on what ideas to include, asked for confirmation on details, and disagreed with proposal content. Yet only 2 of the 8 total marginal comment threads were resolved and led to changes in the proposal, leaving many disagreements, issues, and requests for clarification unaddressed (Figure~\ref{fig:user_timelines}: Merge in Wikum+ vs. Resolve Proposal Comment in Control).

Furthermore, similar to Study 1, the style of writing in the Wikum+ proposal was more concise and consisted of many bulleted lists, likely drawing inspiration directly from the summaries created for subthreads. The Doc proposal, on the other hand, was more essay-like, written in full paragraphs, typically with one or two users responsible for each paragraph. It's possible that users found that making modifications was easier in Wikum+, as opposed to trying to find the right words to fit into an existing paragraph written by someone else. A user in the control group said that they ``\textit{felt really uncomfortable editing others' choice of words since that has a direct impact on how they initially delivered that thought}'', a sentiment that has been observed in other work~\cite{wikum}.

\textbf{Users in Wikum+ presented the final proposal as a full list of options, whereas Doc users went with a majority vote, leading to a lower sense of inclusiveness.} Two distinct summaries were created in the Doc for the first Task point, one in favor of a mandatory Pass or No Record (P/NR) grading scheme and the other in favor of making this grading system opt-in by students. However, a majority of users agreed with the mandatory P/NR idea, resulting in more users contributing to writing that summary. Meanwhile, only 3 users voiced that they preferred the optional P/NR scheme, resulting in an unfinished alternative summary, as fewer users worked on it. After users voted on which summary was preferred, one user in the majority group took the initiative to move the minority summary out of the proposal and to the bottom of the Doc. Furthermore, users in the control began summarizing and writing the proposal very early on---before many users had a chance to provide input and before deliberation on many topics had reached a conclusion (Figure~\ref{fig:user_timelines}). While this set a tone and direction for the proposal, a number of marginal comments that disagreed with already-written proposal text went unresolved.

One user in the Google Doc who rated inclusiveness as low in the post-study survey said: ``\textit{I felt like I was in a dissenting minority. On the topic of grades, the majority group didn't take into account the full extent of pros/cons. Generally, I felt like the doc was written from a perspective that very much was not mine, so it was hard to write much.}'' As a result, some users in the control said that they stuck with ``\textit{edit[ing] things related to grammar, sentence structure, and clarity},'' instead of contributing to writing up an opinion they didn't agree with, reducing their proposal-writing contribution levels (Figure~\ref{fig:user_timelines}).

\begin{figure}
    \centering
    \includegraphics[width=1\columnwidth]{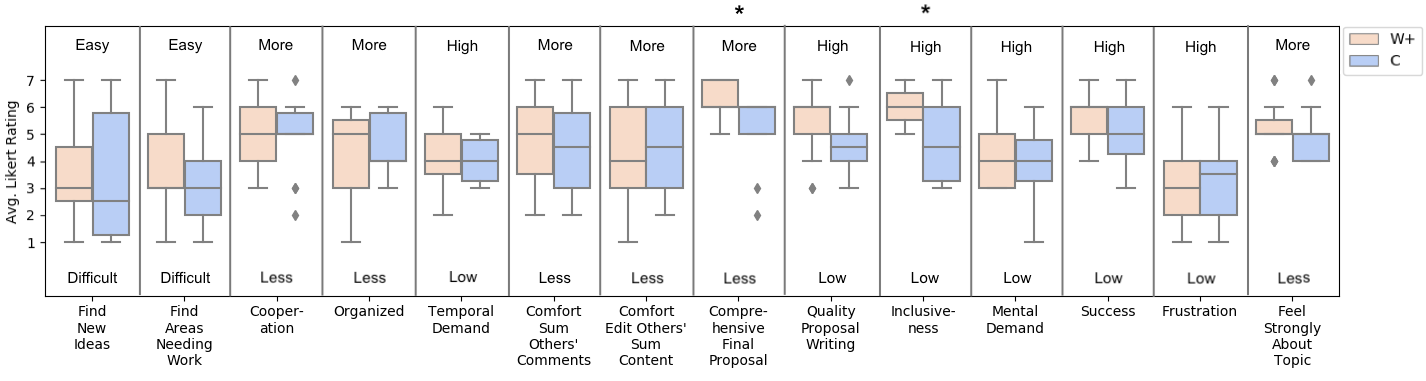}
    \caption{Box plots showing metric values (out of 7), by condition, from the large group conditional post-study surveys. Metrics that resulted in significantly different means from one-way ANOVA tests are starred ( \textbf{*} = $p<0.05$).}
    \label{fig:large_metrics}
\end{figure}

Running a one-way ANOVA test on post-study survey metrics showed a significant difference between the levels of inclusion users experienced in each group ($F=6.12$, $p=0.019$), with Wikum+ users feeling more included than Doc users did (Figure~\ref{fig:large_metrics}). One Wikum+ user said that ``\textit{every comment had responses to it, either agreeing or adding on points, or bringing another viewpoint. They were all discussed, including my comments, so I felt included.}'' 

\textbf{With every side's opinion expressed in Wikum+, more users could collaboratively summarize the discussion, leading to a greater sense of comprehensiveness.} 
Users in the control whose opinions were voted out had a harder time contributing to the final proposal. One user said: ``\textit{The document wasn't finished...there seemed to be a lot of unresolved comments where people voiced disagreement with parts of the first half. Towards the end, though, it seemed mostly like a one to two person effort, so I think this is where the quality of writing falls off...the critiques were not acted upon.}'' With too much time spent on the grading subtopic of the task, coupled with minority voices losing engagement and critiques being unaddressed, the control group's final proposal lacked completeness and the variety of ideas that were earlier deliberated.

Compared to the control group, Wikum+ users felt that their proposal was significantly more comprehensive in content ($F=6.61$, $p=0.016$, Figure~\ref{fig:large_metrics}), with outside evaluations agreeing with this assessment (Table~\ref{tab:large-prop-eval}, Avg. Comprehensiveness -- W+ : 4.5, C: 2.83). The Wikum+ group made use of neutrality, comprehensiveness, and writing quality flags to improve upon summaries cooperatively. As summaries were refined, users updated these flag evaluations from biased to neutral, from poor to great writing quality, and from lacking to complete coverage. Although the results were not statistically significant, participants in Wikum+ indicated a higher level of comfort on summarizing other users' comments, on average, than those in the control (Figure~\ref{fig:large_metrics} -- Wikum+ average comfort: 4.8; control average comfort: 4.4).
One participant said that ``\textit{users took it upon themselves to try and present information in equal light. Later, edits were committed to try and clear up the record and preserve a neutral voice.}'' Another user said that ``\textit{adding comment references and the color scheme of unsummarized comments played a huge role in making sure someone reads everything in the sub-tree when they're summarizing a thread.}'' Although the comments had differences in opinions, the summary synthesizing the comments consisted of pros and cons for multiple options, making the writing less biased (Table~\ref{tab:large-prop-eval}, Avg. Level of Bias -- W+ : 3.33, C: 5.5).
As a result, Wikum+ users incorporated a diverse set of ideas and opinions into the sub-tree summaries,  which were then translated into the final document.

\begin{table}
\centering
\begin{minipage}{\columnwidth}
    \centering
    \begin{tabular}{C{0.7cm}|C{2.4cm}|C{2.4cm}|C{2.4cm}|C{2.2cm}|C{1.4cm}}
      \Xhline{2\arrayrulewidth}
      & \textbf{\footnotesize{Avg. Argument Quality (1: low, 7: high)}} & \textbf{\footnotesize{Avg. Comprehensiveness (1: low, 7: high)}} & \textbf{\footnotesize{Avg. Level of Bias (1: neutral, 7: biased)}} & \textbf{\footnotesize{Avg. Writing Quality (1: low, 7: high)}} & \textbf{\footnotesize{More successful}}\\
      \Xhline{2\arrayrulewidth}
      \footnotesize{W+} &  4 &  4.5 & 3.33 & 3.67 & 4 \\
      \hline
      \footnotesize{C} & 3.67 & 2.83 & 5.5 & 3.5 & 2 \\
    \end{tabular}
\caption{Evaluation of Large Group Proposals. 6 non-participants scored both proposals.}~\label{tab:large-prop-eval}
\end{minipage}
\end{table}

\textbf{Quality of the final proposals.}
As discussed above, the Wikum+ proposal was written in a list-like fashion while the Doc proposal was written in a essay-like style. However, in this study, Wikum+ users found the quality of writing in their proposal higher than Doc users found theirs (Figure~\ref{fig:large_metrics}, Quality of Writing in Proposal -- W+ : 5.3, C: 4.6), though not by a significant amount ($F=2.70$, $p=0.11$). The independent evaluation by external raters similarly indicated that the Wikum+ proposal's writing quality, argument quality, and overall success exceeded that of the control (Table~\ref{tab:large-prop-eval}).
In this study, non-participants also found the Wikum+ proposal to be more comprehensive and less biased than the Doc proposal (Table~\ref{tab:large-prop-eval}), supporting the users' own evaluations of their final documents' comprehensiveness and inclusiveness (Figure~\ref{fig:large_metrics}).


\section{Discussion}

 Wikum+ introduces interleaved discussion and summarization as a workflow for seamlessly switching between the phases of growing a discussion with new ideas and shrinking a discussion through the synthesis of existing ideas. During the studies, we saw users in Wikum+ reaching consensus  thread-by-thread, weaving summarization into the discussion. This allowed for a gradual yet organized process for collectively coalescing numerous ideas into a final proposal. In Study 1, Wikum+ users did not face the same ``activation barrier'' that some control groups had to overcome in collecting everyone's thoughts and moving them to the proposal. In some cases in the control (G3 and G6), one user had to individually take this step for the group---a feat that does not scale to large conversations.

In the studies, we also saw users in Wikum+ iteratively refine summaries by adding post-summary comments to summaries, forming cycles of summarization followed by discussion that then gets merged in.
In Study 1, users felt that Wikum+ helped in finding where new ideas had occurred and in discovering topics still requiring further work. The presentation of post-summary comments in Wikum+ helped users recognize that summaries needed updating to incorporate a new idea or a suggestion for improvement. 
In Study 2, where an increase in scale was reflected in a greater diversity of opinions, we saw how post-summary comments in Wikum+ differed from marginal comments on the Doc proposal in the way ideas were incorporated. In Wikum+, users felt included as neutral summaries were refined to bring in their points-of-view, making for a more comprehensive final proposal representing a wide range of opinions. 
On the other hand, in the control condition, minority opinions were voted out, and marginal comments were left unresolved as users could not agree on how to incorporate the divergent ideas.

At an even larger scale, one could imagine multiple rounds of these summary-discussion cycles taking place, continuously growing and shrinking conversations as summaries are improved upon.
Moreover, as post-summary discussions flourish, one could imagine that recursive loops may emerge where lengthy post-summary discussions grow, are summarized into sub-summaries, and finally, are synthesized back into the original summary, coming full circle. Thus, discussion and summarization need not fall neatly into phases, but can be interchanged continuously, with summaries and deliberation serving multiple purposes. With this process, summaries help users both catch up on the conversation and provide a comprehensive and dynamic synthesis of generated ideas. Discussion not only allows for ideation---it also provides a means for improvement upon generated summaries.

\subsection{Design Implications}
In Study 2, we saw the control group try to converge too quickly on a particular topic within the discussion, stifling further ideation and imposing self-limitations on their resulting proposal, while the Wikum+ group took a more inclusive and comprehensive approach of generating a plethora of ideas and then incorporating all opinions as different options.
However, as it is possible to summarize a subthread at any point in time in Wikum+, one concern may be that users do not spend enough time discussing before jumping to summarize, since introducing convergent thinking too early in the ideation process can hinder creativity~\cite{cropley2016creativity}.
Thus, in Wikum+ we added the capability as the creator of a Wikum+ instance to set the permissions of the instance to be comment-only, edit-only, or both. This way, a group could be forced to participate only by commenting for a period before gaining the ability to summarize and organize the discussion. In doing so, they produce a sufficient quantity of ideas before introducing summarization, which may influence users into a consensus-finding point-of-view. In our lab studies, we did not use this feature, allowing all users a full range of commenting, editing, and summarizing functions, as we wished to observe how groups would naturally act.
However, a balance is required when bringing in summarization---too early and people may begin converging before all ideas are explored, too late and redundancy, off-topic conversations, and information overload could distract and overwhelm users. 
Though we make it a setting that initiators can toggle on a per-instance level to suit each group's needs, future work could examine whether there are any signals, such as the change in velocity of comments, that we can use to suggest to Wikum+ initiators to enable summarization.

In both lab evaluations, Wikum+ groups built their final proposal in a writing style reminiscent of the underlying list-like subtree summaries. While these resulting documents were concise and could be beneficial in some situations, in other cases, users may want their document to read more like an essay. This suggests that a view that users can turn on while writing that looks more like a document might help them get into a document-writing mindset. We could provide such a view in addition to the regular Wikum+ view that combines all the summaries at a user-set level of depth reflecting exactly the content in those summaries. 
For instance, a document could be auto-generated from the subtopic summaries, with each summary as one paragraph in the document. 
Changes in a summary would immediately be reflected in the document, and vice-versa. This integrated document view would ideally help users move towards a more flowing style of writing, as we observed occurred in many of the control condition proposals.

In our post-study surveys, a number of users expressed that a separate ``meta'' discussion space for coordinating efforts and designating tasks would have benefited their group. An existing option for this in Wikum+ is for users to just create a ``meta-thread'' on the page specifically for this type of coordination. However, users may find that this thread does not belong within the discussion itself. Wikum+ could instead designate a collapsible meta-discussion space on the page for arranging and coordinating high-level tasks.

Furthering the notion of summary-discussion cycles, Wikum+ could provide a way to leave comments that annotate a particular part of the summary, thus differentiating post-summary discussion that directly comments on the summary content from more general feedback. 
We currently make post-summary discussion discoverable by reopening summaries and distinguishing incomplete summaries, as well as unincorporated new comments. However, as we saw in our evaluations, post-summary discussions sometimes only addressed one aspect of the summary. A distinguishable annotation on summary text could allow for more fine-grained improvements, encouraging summary-discussion cycles that support more detailed refinement.

\subsection{Broadening an Interleaved Approach to Other Applications and Domains}
 Wikum+ introduces processes adaptable to other software platforms and other forms of collaboration beyond proposal writing. In Study 1, we saw how all three groups that had the control condition second replicated strategies inherent in the Wikum+ interface that the groups had used in the previous round. One user described their second round control condition: ``\textit{We learned to break up different sections of the report and work on those sections and then eventually take those comments, summarize them, and put them all back together into one essay}.''
From the organized synthesis process of threaded discussion to interleaving summarization and deliberation, our system contributes a workflow reproducible on various other discussion interfaces. Reflecting Wikum+, a natural enhancement of threads on Reddit could embed the ability to create distinct mutable summaries from comments. Similarly, Wikipedia talk pages could link a discussion to the resulting synthesis on the final page, allowing for continuous, traceable enhancement. As seen from our lab studies, the benefits of Wikum+ in increasing inclusiveness, comprehensiveness, and document quality are enhanced in larger group sizes. Thus, while smaller groups could hold a call or a meeting to deliberate, Wikum+ may prove advantageous when there are more participants, ideas, and conversation.

In our lab evaluations, we studied the case of collaboratively creating a proposal based on a collection of ideas. But other types of discussions could also use Wikum+ to serve a multitude of goals. For instance, Wikum+ can be repurposed to enhance the well-structured interface of The Deliberatorium built for argumentation~\cite{klein2011harvest} or provide a platform for contributors in the Knowledge Accelerator~\cite{hahn2016knowledge} to discuss credibility of crowd-sourced information and improve upon the resulting article. Similarly, in the realms of visual, audio, or video collaborations, interleaving synthesis into ideation could help groups organize and collect thoughts, while gradually refining and  iteratively improving their resulting product. 

More broadly, researchers in creativity support and UX and product design have focused on the design process as comprised of set phases for  divergent thinking---where diverse ideas or solutions are generated---and convergent thinking---where those ideas are synthesized and optimal ideas are identified \cite{bobbe2016comparison}. These design processes, including the standard ``Double Diamond'' approach in design work \cite{council2005study}, and comparable models with more iterative feedback loops \cite{roozenburg1995product} have similar variants across many disciplines \cite{gericke2012analysis}. However, research shows that these models are often too simple and idealistic \cite{parnas1986rational}, poorly representing actual more sophisticated creative processes \cite{howard2008describing}, where feedback and new ideation can continue during the convergence process.
Wikum+ provides an interface to join divergent and convergent thinking within the same space, making the transition between the two processes far more fluid and reducing the barrier for beginning the synthesis process. This allows for smoother and more connected transitions between the two phases, and allows for continuous feedback and improvements through summary-discussion cycles.



\section{Future Work and Limitations}
In our user studies, a large proportion of the participants were female. While one of the smaller mailing lists we recruited from in Study 1 was all-female, the bulk of our recruiting was from general university mailing lists consisting of a diverse undergraduate population. The first forty undergraduate students to sign up in Study 1 were selected and everyone that expressed interest in Study 2 was selected. Thus, the gender ratios in our studies were likely self-selecting, but ended up in a female majority.

Furthermore, Study 2 only consisted of one group in each condition. However, group dynamics can vary greatly~\cite{noel2004empirical}, and the participants in each group may not be acting independently, as users in a group can influence the behaviors and actions of others. Therefore, further studies with more groups are needed to support the evidence we found in our work. Additionally, because we evaluated Wikum+ using lab studies, further work is needed to understand how collaborators would use our system in the wild. In particular, more controversial topics, the presence of online trolls, and a larger scale of users may affect how Wikum+ performs. We note that the tool does allow initiators to block individuals from actions; however, it's unclear how bad actors could game the system in a real setting.

In our design and evaluation of Wikum+, we focused on threaded discussion structures, which naturally provide a powerful grouping heuristic. Summarization embedded in deliberation could be extended to other threaded discussion interfaces as well. However, future work should also study the effects of similar interleaved discussion and summarization processes in non-threaded discussions, such as in instant messaging conversations. While tools for summarization in platforms such as Slack exist \cite{zhang2018making}, evaluation is needed to study how users may use similar synthesizing methods as in Wikum+ to collaboratively create a document from their conversation directly in the tool. As discussed previously, our work focuses on textual collaboration, but further research is needed to study the effects of interleaved ideation and synthesis in other collaborative mediums such as in visual, audio, and video work.

Finally, data collected from summarization on Wikum+ could provide training data to improve and build upon NLP models for automatic summarization of discussions. Wikum+ could also automatically generate initial higher-level summary suggestions from the contents of intermediate summaries for users to reference and modify.

\section{Conclusion}
In this work, we designed, developed, and evaluated a process for interleaved discussion and summarization and a system called Wikum+ that allows groups to collaboratively create a living summary artifact directly through the synthesis of their discussions. By interchanging discussion, where new ideas and feedback grow the conversation, and summarization, where consensus and synthesis shrink the conversation, resulting documents can be continuously refined. From our lab studies, we found that Wikum+ allowed for greater organization and light-weight coordination in smaller groups, and more inclusion and comprehensiveness of the resulting document in larger groups.



\bibliographystyle{ACM-Reference-Format}
\bibliography{sample-base}

\received{June 2020}
\received[accepted]{July 2020}

\appendix
\section{Appendix: Proposal Tasks}\label{appa}

The following tasks were given to groups in our lab evaluations. Wikum+ users were instructed to leave their final proposal in the top level summaries of the page while users in the control condition were instructed to write their proposal in the first few pages of the Google Doc. \\

\noindent \textbf{Quarantine Online Classes Task}

\noindent Your task is to come up with a proposal on how online classes should operate next semester if students are not allowed to return to campus.
This proposal should include a number of components:

\begin{enumerate}
    \item How the grading system should be, including your group’s opinions on the pros and cons of a mandatory Pass or No Record grading scheme
    \item Whether tuition should or should not be reduced and why
    \item How much homework workload should be given
    \item How midterms and finals should be structured
    \item How student resources should be provided
\end{enumerate}

 \vspace{2mm}
\noindent \textbf{Campus Dining Options Task}

\noindent Your task is to come up with a proposal on how the school administration can improve dining options around campus. This proposal should include a number of components:

\begin{enumerate}
    \item Why current dining options are problematic
    \item A comprehensive list of dining solutions
    \item A cohesive and convincing argument of best 1-2 options from list
    \item A trade-off analysis of the selected options
    \item A proposed timeline on how to implement the selected options by 2020
\end{enumerate}
\end{document}